\documentclass{aa}
\pdfoutput=1
\usepackage[pdftex]{color,xcolor}
\usepackage[pdftex]{graphicx}
\usepackage{wasysym}
\usepackage{hyperref}
\usepackage{txfonts}

\def\mearth{M_\oplus}
\def\msun{M_\odot}

\def\mcore{M_{\rm core}}

\def\astart{a_{\rm start}}
\def\fpg{f_{\rm D/G}} 
\def\fpgsun{f_{\rm D/G,\odot}} 

\def\f1{f_{\rm I}}
\def\mj{M_{\textrm{\jupiter }}}
\def\mstar{M_*}
\def\mdisk{M_{\rm disk}}
\def\tdisk{\tau_{\rm disk}}
\def\astart{a_{\rm start}}
\def\tstart{t_{\rm start}}

\def\anorm{a_0}

\def\mdotcore{\dot{M}_{\rm core}}
\def\mwind{\dot{M}_{\rm w}}

\def\aice{a_{\rm ice}}
\def\beq{\begin{equation}}
\def\eeq{\end{equation}}
\def\mplanet{M_{\rm planet}}

\def\sigmas{\Sigma_{\rm S}}
\def\sigmassn{\Sigma_{\rm S,SN}}

\def\sigmanorm{\Sigma_{0}}

\def\msini{M\sin i}
\def\amax{a_{\rm max}}
\def\amin{a_{\rm min}}

\def\mm{[M_{\rm D}/M_{\rm SN}]}

\def\aj{AJ}                   
\def\araa{ARA\&A}             
\def\apj{ApJ}                 
\def\apjl{ApJ}                
\def\apjs{ApJS}               
\def\ao{Appl.Optics}          
\def\aap{A\&A}                
\def\mnras{MNRAS}             
\def\pasp{PASP}               

\def\icarus{{Icarus}}

\def\({\left(}
\def\){\right)}
\def\<{\left<}
\def\>{\right>}

\begin{document}

\title{Extrasolar planet population synthesis } 
\subtitle{IV. Correlations with disk metallicity, mass and lifetime}

\author{C. Mordasini\inst{1}  \and Y. Alibert\inst{2,3}  \and W. Benz\inst{2}  \and H. Klahr\inst{1} \and T. Henning\inst{1}}

\institute{ Max-Planck-Institut f\"ur Astronomie, K\"onigstuhl 17, D-69117 Heidelberg, Germany \and
Physikalisches Institut, University of Bern, Sidlerstrasse 5, CH-3012 Bern, Switzerland \and  
Institut UTINAM, CNRS-UMR 6213, Observatoire de Besan\c{c}on, BP 1615, 25010 Besan\c{c}on Cedex, France } 

\offprints{Christoph MORDASINI, \email{mordasini@mpia.de}}

\date{Received 27.05.2011 / Accepted 27.12.2011}

\abstract{This is the fourth paper in a series showing the results of planet population synthesis calculations. In Paper I, we presented our methods. In Paper II, we compared statistically the synthetic and the observed planetary population. Paper  III addressed the influences of the stellar mass on the population.}
{Our goal in this fourth paper is to systematically study the effects of important disk properties, namely disk metallicity, mass and lifetime on fundamental properties of planets like mass and semimajor axis.}
{For a large number of protoplanetary disks which have properties following distributions derived from observations, we calculate a population of planets with our formation model. The model is based on the classical core accretion paradigm but includes self-consistently  planet migration and disk evolution.}
{We find a very large number of correlations: Regarding the planetary initial mass function, metallicity, $\mdisk$ and $\tdisk$ have different roles: For high metallicities, giant planets are more frequent. For high $\mdisk$, giant planets are more massive. For long $\tdisk$, giant planets are both more frequent and massive. At low metallicities, very massive giant planets cannot form, but otherwise giant planet mass and metallicity are nearly uncorrelated. In contrast, (maximal) planet masses and disk gas masses are correlated. The formation of giant planets is possible for initial planetesimal surface densities $\sigmas$ of at least 6 g/cm$^{2}$ at 5.2 AU. The sweet spot for giant planet formation is at $\sim5$ AU. In- and outside this distance, higher $\sigmas$ are necessary. Low metallicities can be compensated by high $\mdisk$, and vice versa, but not ad infinitum. At low metallicities, giant planets only form outside the ice line, while at high metallicities, giant planet formation occurs throughout the disk. The extent of migration increases with $\mdisk$ and $\tdisk$ and usually decreases with metallicity. No clear correlation of metallicity and the semimajor axis distribution of giant planets exists because in low metallicity disks, planets start further out, but migrate more, while the contrary applies for high metallicities. The final semimajor axis distribution contains an imprint of the ice line. Close-in low mass planets have a lower mean metallicity than Hot Jupiters. The frequency of giant planets varies approximately as $\mdisk^{1.2}$ and $\tdisk^{2}$.}
{The properties of protoplanetary disks - the initial and boundary conditions for planet formation - are decisive for the properties of planets, and leave many imprints in the population.}


\titlerunning{Extrasolar planet population synthesis IV}

\authorrunning{C. Mordasini et al.}

\maketitle

\section{Introduction}\label{sect:introduction}
The number of known extrasolar planets has grown large enough to look at the statistical properties of the population as a whole, rather than at the properties of single objects, and to compare the actual population with a synthetic population obtained from a theoretical planet formation model. In this way \textit{all} discovered planets serve (provided the detection bias is known)  to constrain the model, and to improve our understanding of planet formation. We used our extended core accretion formation model (Alibert et al. \cite{alibertetal2005a}) to generate in a Monte Carlo way populations of synthetic extrasolar planets (Mordasini et al. \cite{mordasinietal2009a}, hereafter Paper I). Then, we compared the detectable synthetic planets with an observational comparison sample of actual exoplanets using statistical methods (Mordasini et al. \cite{mordasinietal2009b}, hereafter Paper II). We found that we could reproduce in a statistically significant way some of the most important  properties of the observed extrasolar giant planets. Finally, in  Alibert et al. (\cite{alibertetal2010}), hereafter Paper III, we discussed the influence of the stellar mass on the synthetic planetary population, studying for example the effect on the planetary initial mass function, the semimajor axis distribution or the ``metallicity effect''  (the increase of the detection probability of giant planets with metallicity). In this Paper IV, we focus back onto solar-like stars. 

Over the last years, observational considerable progress was achieved in the characterization of the end products of planetary formation process, i.e. the planet themselves. Progress has also been substantial in the characterization of the initial and boundary conditions for this process, i.e. the properties of protoplanetary disks. This was made possible  to a large measure thanks to new observational facilities like \textit{Spitzer} (e.g. Fang et al. \cite{fangetal2009}). Observations of disks around young stars have provided us with knowledge of the distributions of disk masses (Beckwith \& Sargent \cite{beckwithsargent1996}; Andrews et al. \cite{andrewsetal2009}),  disk sizes (McCaughrean \& O’dell \cite{mccaughreanodell1996}; Andrews et al. \cite{andrewsetal2010}) and lifetimes (Haisch et al. \cite{haischetal2001};  Fedele et al. \cite{fedeleetal2010}). Additionally, correlations between disk properties and stellar mass were discovered (Kennedy \& Kenyon \cite{kennedykenyon2009}; Mamajek \cite{mamajek2009}), which have important implications for the formation of planets around stars of different masses (Currie \cite{currie2009}, Paper III). As was shown in Paper I, the diversity of extrasolar planets is a direct consequence of the diversity of protoplanetary disks, which means that the properties of disks are critical in defining the outcome of planet formation taking place in such disks.

In this article we focus on correlations between disk and planetary properties. We study systematically the influences of  disk metallicity [Fe/H], disk (gas) mass $\mdisk$ and lifetime $\tdisk$ on important, observable properties of synthetic planets by computing a large number of models of planet formation. 

\subsection{Observed correlations}
From an observational point of view, a number of correlations were inferred in the past years, with various degrees of significance (see Udry \& Santos \cite{udrysantos2007} or Mayor et al. (\cite{mayormarmier2011}) for an overview). We address in this paper the following correlations: 

(1) The clearest correlation is the link between the stellar metallicity (to first order an indicator of the initial dust-to-gas-ratio in the disk, see Santos et al. \cite{santosetal2003}) and the likelihood of detecting a giant planet. This ``metallicity effect'' is observationally known for a long time and very well established for solar-like stars by numerous studies  (e.g. Gonzalez \cite{gonzalez1997}; Fischer \& Valenti \cite{fischervalenti2005}, Udry \& Santos \cite{udrysantos2007}).  The ``metallicity effect'' was studied as an observational constraint in Paper II and is here further addressed in Sect. \ref{sect:imfobservabledist}, \ref{maximalmassesandmetallicity} and \ref{sect:detectprobmetallicity}.  

(2) Interestingly, lower mass Neptunian and Super-Earth planets seem in contrast not to be found preferentially around high [Fe/H] stars (Mayor \& Udry \cite{mayorudry2008}; Sousa et al. \cite{sousaetal2008}; Ghezzi et al. \cite{ghezzietal2010}).  We study this in Sect. \ref{subsubsect:imf} and \ref{sect:metallicitycloseinplanets}.

(3) Already much less compelling is a possible absence of  very massive planets (in a mass range of about 5 to 20 Jupiter masses, $\mj$) orbiting within a few AUs low metallicity solar-like stars as found in radial velocity (RV) searches. Such a possible absence was noted by Udry et al. (\cite{udryetal2002}), Santos et al. (\cite{santosetal2003}) and Fischer \& Valenti (\cite{fischervalenti2005}). This is discussed in Sect. \ref{sect:compobsmmaxfeh}.

(4) In contrast, no secure correlations were found between stellar metallicity and planetary semimajor axis, at least among RV detections (Udry \& Santos \cite{udrysantos2007}; Valenti \& Fischer \cite{valentifischer2008}; Ammler-von Eiff et al. \cite{ammlervoneiffetal2009}), even though possible correlations were discussed in the literature. Sozzetti (\cite{sozzetti2004})  for example studied whether stars with Pegasi planets have particularly high [Fe/H], even among planet hosts. Correlations between migration, semimajor axis and metallicity are addressed  in Sect. \ref{sect:extentofmigration} and \ref{sect:semimajoraxisdist}. 

(5) Giant planets are  more  frequently found orbiting A stars than orbiting solar-like stars, and are also more massive (Lovis \& Mayor \cite{lovismayor2007}; Bowler et al. \cite{bowleretal2010}) on average. In Paper III it was shown that this correlation can be best reproduced if one assumes a roughly linear scaling between disk (gas) mass and stellar mass. This implies, at least partially, a correlation between disk mass and planet mass (Sect. \ref{sect:pimfdiskgasmass} and \ref{sect:maximalmassesanddiskmass}).   

In addition, we address below several other correlations that appear in the models but that have not yet been reported in the observation, but which could become observable in the future with better instruments and larger, more complete data bases. 

\subsection{Special role of [Fe/H]}
Population synthesis is a powerful tool to investigate such correlations, because all properties of the (numerical) parent disks, and those of the synthetic planet populations emerging from them are known. For actual exoplanets this is obviously not the case. Observationally, the host stars metallicity only can be determined and assumed to be a proxy for the disks dust-to-gas ratio $\fpg$. Other fundamental parameters of the disks, such as their initial gas mass (which together with $\fpg$ sets the absolute amount of solids), and their lifetime, cannot be observationally determined. It is therefore no surprise that the observationally inferred correlations involve essentially only the metallicity. 

A recently discussed correlation that could partially change this is the possible enhanced lithium depletion measured in Sun-like stars hosting planets (Israelian et al. \cite{israelianetal2009}, but see Baumann et al. (\cite{baumannetal2010}) for an opposed view). This is because such a depletion could result from a positive correlation of disk lifetime and likelihood of forming giant planets (Bouvier \cite{bouvier2008}).  The influence of  the disk lifetime $\tdisk$ on the occurrence of giant planets, as observed in our models,  is discussed in Sect. \ref{sect:detectprobtdisk}. 

In general however, only probability distributions for disk masses and lifetimes can be inferred from the observations of star forming regions. Our study allows to a certain degree to narrow down these probability distributions for a specific star-planet system, as not all types of planets can form in all types of disks. 

\subsection{Earlier works}
Several studies have already addressed correlations between disk and planetary properties:  Ida \& Lin (\cite{idalin2004b}) studied the influences of the metallicity as modeled by the dust-to-gas ratio. They find that the normalized mass and semimajor axis distributions of the potentially observable (giant) synthetic planets are rather independent of [Fe/H], while the frequency of giant planets increases with it, in agreement with observations.  Kornet et al. (\cite{kornetetal2005}) modeled the evolution of the solids from dust size to planetesimals and, using the final planetesimal surface density, estimated the giant planet formation capability. Such an approach also reproduces the ``metallicity effect''.  Dodson-Robinson et al. (\cite{dodsonrobinsonetal2006}) derived a fit for the time until gas runaway accretion is triggered by a planetary core at a fixed distance of 5.2 AU, as a function of the planetesimal surface density.  They used this fit to estimate the frequency of giant planets as a function of several disk properties. We show our results regarding this subject in Sect. \ref{sect:detectprob}, comparing with their results. Matsuo et al. (\cite{matsuoetal2007}) used  parameterized  formation models to determine the disk parameters where giant planet formation is possible either by core accretion or direct gravitational collapse. We present a similar study in Sect. \ref{sect:diskcondleadingtogiants}.

\subsection{Structure of the paper}
The paper is organized as follows: Section \ref{sect:methods} describes the methods used to obtain the result concerning the synthetic population presented in Sect. \ref{sect:amdiagram}. Section \ref{sect:mass} studies various aspects of the impact of disk properties on the mass of extrasolar planets, including the planetary initial mass function.  Section \ref{sect:diskcondleadingtogiants} analyzes under which disk conditions giant planets can form. Section  \ref{sect:semimajoraxis} addresses the correlations of disk properties with planetary migration and the final semimajor axis distribution. In Sect. \ref{sect:metallicitycloseinplanets} we study the metallicity of close-in planets. Section \ref{sect:detectprob} shows how disk properties determine the fraction of stars with giant planets, while Sect. \ref{sect:influencesdisklifetime} asses the consequences if disk masses and lifetimes are correlated. Finally, in Sect. \ref{summaryconclusion} we summarize the results and present our conclusions. 

\section{Methods}\label{sect:methods}
As our approach to planet formation and population synthesis was described in details in Paper I, we limit here ourselves to a short overview.

\subsection{General Procedure}
To obtain a synthetic population of planets, we proceed in five steps: (1) The probability distributions for the initial conditions are derived from observations of protoplanetary disks. (2) A large numbers of sets of initial conditions are drawn from these probability distributions in a Monte Carlo fashion. (3) The corresponding final outcomes of the planet formation process (planetary mass and position) are computed using our planet formation model. This results in a population of synthetic planets, most of them however undetectable with current observational techniques.  (4) To obtain the subset of the potentially observable synthetic planets, we apply an appropriate detection bias. Since most planets have been discovered by radial velocity techniques, we use a bias based on the velocity amplitude of the host star.  (5)  Correlations between the initial (disk) conditions and the planetary properties are searched and compared to observations (if existing).

\subsection{Initial Conditions - Probability distributions}
We use four Monte Carlo variables to specify the initial conditions.  (1) The dust-to-gas-ratio $\fpg$ which we link to [Fe/H] as [Fe/H]=$\log(\fpg/\fpgsun)$, where $\fpgsun$ is the dust-to-gas ratio of the solar nebula for which we assume a value of 0.04. The choice of this value was discussed in details in Paper I. Here we only briefly mention that the factor two to three by which 0.04 is larger than the measured photospheric Z of the sun (Lodders \cite{lodders2003}) is a first order representation of the effects of dust evolution and drift. These mechanisms lead to an increase of the ``planetesimal'' $\fpg$ compared to the original ``dust'' $\fpg$  in the inner, planet forming parts of the disk by a similar factor through the advection of material from the outer disk (Kornet et al. \cite{kornetetal2004}). The probability of occurrence of a given [Fe/H] is derived from the metallicity distribution of the FGK stars in the CORALIE  planet search sample (Udry et al. \cite{udryetal2000}) which is representative for the [Fe/H] distribution of the target stars in the various major radial velocity search campaigns (Paper I).  Thus, we assume that the observed stellar metallicity is a good indicator of the primordial disk metallicity (Santos et al. \cite{santosetal2003}, but see also Pasquini \cite{pasquinietal2007}). (2) The initial gas surface density $\sigmanorm$ at $\anorm=5.2$ AU which we link to the initial disk gas mass $\mdisk=4 \pi \sigmanorm \anorm^{3/2}(\sqrt{\amax}-\sqrt{\amin}) $.  The probability distribution of circumstellar disk masses is derived from the observations of the $\rho$ Ophiuchi star formating region (Beckwith \& Sargent \cite{beckwithsargent1996}). (3) The rate of photo-evaporation $\mwind$.  The distribution of $\mwind$ is constrained by the observed disk age distribution: We have adjusted our $\mwind$ distribution to obtain together with our value of the viscosity parameter $\alpha$ a distribution of disk lifetimes  $\tdisk$ that is in good agreement with the observed distribution (Haisch et al. \cite{haischetal2001}).  (4) The initial semimajor axis of the planetary seed $\astart$. The distribution of the starting positions of planetary embryos is not constrained by observations and only theoretical arguments can be used. They indicate (e.g. Mordasini et al. \cite{mordasinietal2009a}) that runaway bodies should emerge with a uniform distribution in log($\astart$), as already adopted by Ida \& Lin (\cite{idalin2004a}). 

In addition, we assume that these Monte Carlo variables are  independent variables.  Hence, we do not take into account potential correlations arising during disk formation itself e.g. via opacity effects.  The fact that we draw $\sigmanorm$ and $\mwind$ independently implies, on average, longer disk lifetime $\tdisk$  for more massive disks. The importance of this coupling is discussed in sect. \ref{sect:influencesdisklifetime}. We do not consider here the influence of varying initial disk radii (Kornet et al. \cite{kornetetal2005}).  This will be considered in future work, taking into account recent observational results (Andrews et al. \cite{andrewsetal2009,andrewsetal2010}).
 
\subsection{Planet Formation Model}
We use a slightly modified version (see Mordasini et al. \cite{mordasinietal2009a}) of the extended core accretion formation model described in details in Alibert et al. (\cite{alibertetal2005a}). As in Pollack et al. (\cite{pollacketal1996}), we compute the evolution of the planetary core and envelope structure, but include disk evolution using the $\alpha$ formalism, and planetary migration (isothermal type I and type II).  We have been able to show (Alibert et al. \cite{alibertetal2005b}) that the model reproduces many observational constraints imposed by our own giant planets. 

\subsubsection{Relevant model assumptions}\label{sect:relevantmodelassumptions}
For the disk-planet correlations discussed in this work, a number of model assumptions  were found to be particularly relevant and have directly visible consequences in the correlations. 

The first one concerns the structure of the protoplanetary disk, and specifically the location of the iceline.  In the nominal model, the position of the ice line is an increasing function of disk mass due to viscous dissipation (Paper I and III; Min et al. 2011). It is found that this has important consequences both on the mass of giant planets (Sect. \ref{sect:effectcausedbyaice}) and their formation location (Sect. \ref{sect:anticorrfehastart}).

A second assumption which is directly relevant for the masses of giant planets (Sect. \ref{sect:verymassiveplanets}, \ref{sect:maximalmassesanddiskmass}) is that we assume that gap formation does no reduce the gas accretion rate of giant planets (Lubow et al. \cite{lubowetal1999}). As explained in Paper I, the reason for this is that Kley \& Dirksen (\cite{kleydirksen2006}) have shown that when a giant planet becomes sufficiently massive, the disk-planet system can undergo an eccentric instability. The planet then leaves the clean parts of the gap, resulting in a substantial increase of the gas accretion rate. This means that the gas accretion rate is obtained for low mass planets ($\mcore\lesssim 10 \mearth$) by solving the planetary structure equations   while for more massive planets in the disk limited, runaway gas accretion phase it is equal to the  rate at which gas viscously flows  towards the star in the disk $\dot{M}_{\rm enve}=\dot{M}_{\rm disk}=3 \pi \nu \Sigma$ where $\nu$ is the disk viscosity, and $\Sigma$ the gas surface density. 

A third setting which is relevant for the correlation of migration and metallicity  (\ref{sect:collectioneffect}) is that we assume that the planetesimal accretion rates are independent of the migration rate.  The $\mdotcore$ is calculated in the same way as in Pollack et al. (\cite{pollacketal1996}). Possible shepherding effects are not included for the reasons explained in Paper I. This means that cores in low solid surface density disks still can grow relatively massive by migration.

Fourth, we assume for type II migration that as soon as the planet is more massive than the local disk mass ($M_{\rm planet}>\Sigma a^{2}$ where $a$ is the semimajor axis), the migration rate is given as $-(3 \nu/ a)  \times (\Sigma a^{2}/M_{\rm planet})^{p}$. In the nominal model, we use $p=1$ (``fully suppressed'' case,  Armitage \cite{armitage2007}). This is important to understand the absence of a strong correlation of [Fe/H] and the semimajor axis (Sect. \ref{sect:brakingeffect}, \ref{sect:independencefromfeh})

Note that we simplify the problem significantly by assuming that only a single planet can form in a given disk. This would be correct in the limit that proto-planets can form within a single system without influencing each other. This is clearly an idealization, but in the context of this study this is not necessarily a disadvantage. It allows one to see clearly disk-planet correlations, which otherwise might get partially blurred due to the random character of the (gravitational) interactions between several protoplanets. Thommes et al. (\cite{thommesetal2008}) discuss extensively some of the effects induced by the concurrent formation of several planets in one disk. 

The population discussed here is essentially obtained, except if otherwise mentioned, with identical parameters and Monte Carlo distributions as the nominal population presented in Paper I and II.  In particular, this means that the stellar mass is equal to one solar mass $\msun$, the gas disk viscosity parameter $\alpha$ is 0.007, and the type I migration efficiency factor $\f1$ is 0.001. 

However, there is one relevant aspect in which the procedure used here differs from the one used in paper I and II.  We draw the initial semimajor axis of the starting seed strictly uniform in $\log(\astart)$ and disregard the additional criteria mentioned in Paper I. This eliminates some correlations present already in the initial conditions and thus makes it easier to identify the influence of disk properties on the planetary properties. 

As a practical unit for the disk gas masses we define, in analogy to [Fe/H], a relative logarithmic unit, denoted $\mm=\log{(\sigmanorm / \Sigma_{\rm 0,SN})}$, where we assume an initial gas surface density at 5.2 AU $\Sigma_{\rm 0,SN}=200$ g/cm$^{2}$. This is somewhat more than a plain MMSN values  - Hayashi's (\cite{hayashi1981}) value would be about 145 g/cm$^{2}$ -  but with this choice $\mm$ covers, similar as [Fe/H], a nearly symmetric range around zero between -0.6 and +0.7 for the values of $\sigmanorm$ considered here (50-1000 g/cm$^{2}$, corresponding to disk masses between about 0.004 and 0.09 $\msun$). As the disk gas mass $M_{\rm D}$ and the initial gas surface density $\sigmanorm$ are directly proportional to each other, these two terms are often used in an interchangeable way.

\section{Mass-Distance Diagram}\label{sect:amdiagram}
\begin{figure}
\includegraphics[width=\columnwidth]{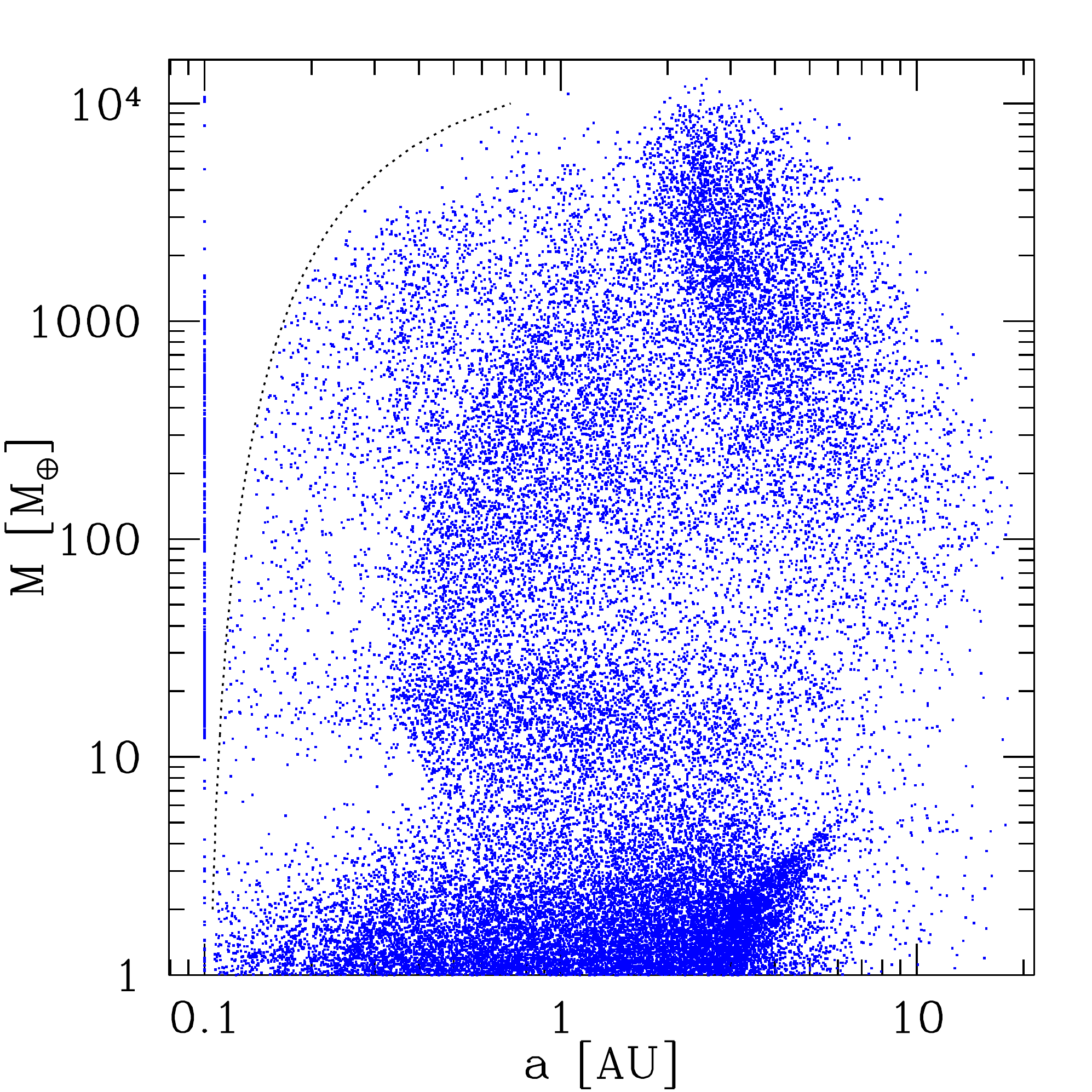}
\caption{Mass-distance diagram of the synthetic population analyzed in this work. The dashed curve shows the feeding limit (cf. Paper I): planets reaching this limit have been arbitrarily set to 0.1 AU.}\label{fig:aM4p}
\end{figure}

To provide an overview of the synthetic planet population we analyze in this work we plot in Fig. \ref{fig:aM4p} the mass-distance diagram of the entire population. In the figure, a number of structures already discussed in Paper I can be identified again, like the low mass ``failed cores'', the approximately Neptunian mass ``horizontal branch'' or the massive ``outer group'' planets outside a few AU. We also see that there are no very massive planets close-in. This is a consequence  of the fact that planets with a mass larger  than the local disk mass migrate at a reduced rate because their inertia is too large for the finite angular momentum flux in the disk. Their absence at small $a$ depends upon the assumed degree of reduction of the type II migration rate once the planet is massive (Syer \& Clarke \cite{syerclarke1995}).

\subsection{Paucity of low mass, close-in planets}
There is also a relative paucity of low mass ($4\lesssim M\lesssim10 \mearth$) close-in ($a\lesssim 0.3$ AU) planets. Recent observations (e.g. Howard et al. \cite{howardetal2010}; Mayor et al. \cite{mayormarmier2011}) rather indicate a high frequency of low mass Super Earth planets close to their parent star.

Concerning this issue, we first remind that our model shows the state of (proto-)planets at the moment when the disk disappears, and not the final one after billions of years of evolution. At this stage, the low mass protoplanets at small distances have a mass about equal the local (solid) isolation mass (type I migration is strongly reduced in this simulation), which is of order  0.1 $\mearth$ only. Further growth from these masses to the final masses by giant impacts (e.g. Marcus et al. \cite{marcusetal2010}) will then only occur after the damping influence of the gas disk is gone (Ida \& Lin \cite{idalin2010}), a phase which is however not included in our model at the moment. This effect would populate the depleted region from ``below''.

\begin{figure*}
\centering 
\includegraphics[width=\textwidth]{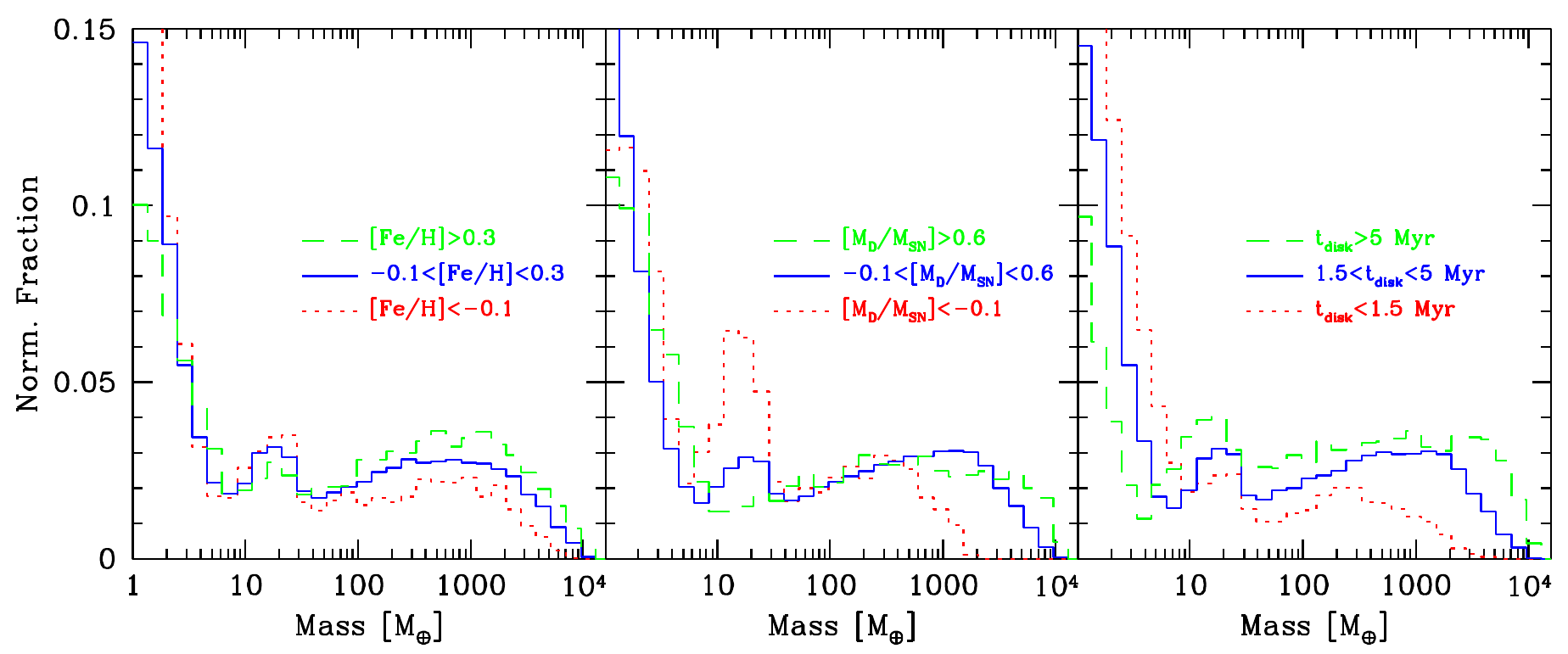}
 \caption{Planetary initial mass function at the moment when the gaseous disk disappears. Note that the models doesn't include any growth processes like giant impacts after disk dispersal which can significantly modify the mass distribution afterwards, especially for low mass planets with $M\lesssim 10 \mearth$. Left: PIMF binned according to metallicity.  Center: PIMF binned according to the gas disk mass. Left: PIMF binned according to the disk lifetime. The meaning of the different lines, and the limiting values, are indicated in the plot.}\label{fig:imfall} \end{figure*}

Second, the high observed frequency could be indicative of quite efficient type I migration, at least in some parts of the disk. This would populate the depleted region from ``outside''. Large quantities of low mass, close-in planets are found in some synthetic populations too, but only for high type I efficiency factors ($\f1\sim0.1-1$) as shown in Fig. 11 of paper II.   On the other hand can the distribution of semimajor axes of giant planets at larger distances only be reproduced when the type I migration rate (as found from linear theories for isothermal disks) is reduced by a significant factor ($\f1\lesssim0.01$, Paper II; Schlaufman et al. \cite{schlaufmanetal2009}). Here we are using the same low type I efficiency factor $\f1=0.001$ which lead in Paper II to the best reproduction of the observed properties of giant planets. Unfortunately, this causes in the same time necessarily also  the depleted region. 

The answer to this dilemma is probably a significantly more complex migration pattern  than what can be mimicked with global (independent of planet mass and distance) efficiency factors like $\f1$  for isothermal type I rates as done here. Recent studies of type I migration dropping the often inappropriate assumption of isothermality (e.g. Paardekooper et al. \cite{paardekooperetal2010}; Masset \& Casoli  \cite{massetcasoli2010}) indeed find complex patterns with rapid in- and outward migration. First exploratory planet population syntheses using such updated type I migration models (Mordasini et al. \cite{mordasinietal2011})  find that type I migration is directed outward in some parts of the disk, and inward in others, which leads to the existence of convergence zones (migration traps) which can result in a pile-up of many low mass planets in certain parts of the disk (cf. Lyra et al. \cite{lyraetal2010}; Sandor et al. \cite{sandoretal2011}).  This important subject will be addressed in dedicated work (Dittkrist et al. in prep.).   

A third mechanism occurs when several planets form and migrate concurrently. Then, lower mass planets can be pushed close to the star after being captured in mean motion resonances of a more massive, more rapidly migrating outer planet. This will be studied in oncoming simulations allowing the formation of many planets per disk (Alibert et al. in prep.)

\section{Mass}\label{sect:mass}
\subsection{Planetary initial mass function (PIMF)}\label{subsubsect:imf}

A central outcome of population synthesis is the planetary initial mass function PIMF (Paper II and III). Figure \ref{fig:imfall} shows the PIMF of all synthetic planets with a mass larger than 1 $\mearth$, binned into low, medium and high metallicity (left panel), disk gas mass (central panel) and disk lifetime (right panel). The diving values are indicated in the plot and are chosen for all three cases in a way that the central bin contains about 70\% of the planets, and the other two about 15\% each. We recall that our models start with an initial seed mass of 0.6 $\mearth$, so that we cannot reliably make predictions about the exact form of the PIMF in the $\lesssim 10\, \mearth$ domain as mentioned in Paper I and II. Note that each bin has been normalized individually, so that the absolute height of different bins (e.g. low vs. medium metallicity) cannot be compared in absolute terms, but taking into account that the initial distribution of [Fe/H] follows the distribution observed in the solar neighborhood. However, one can directly see the relative importance of a given planetary type within a bin (e.g. at high metallicity there are more Jovian than Neptunian planets, in contrast to the low [Fe/H] case.). One recognizes again a number of features discussed in Paper I, like the high mass tail in the Super-Jupiter domain, the giant's plateau in the Jovian mass regime, the minimum at about 30-40 $\mearth$ corresponding to the planetary desert (Ida \& Lin \cite{idalin2004a}), the Neptunian bump and the strong rise towards small masses.  

\subsubsection{PIMF as function of metallicity}
The left panel shows that the metallicity just modifies the importance of the three basic families of planets visible in the PIMF (Jovian planets, Neptunian planets, proto-terrestrial planets), without however changing the general shape of the PIMF (in contrast to the other two disk properties). For giant planets this means that in a high metallicity environment, more giant planets form, but their mass distribution is similar.  The reason for this is related to the fact that a certain [Fe/H] acts as a threshold  for giant planet formation (threshold solid surface density to reach a critical core mass, see Sect. \ref{sect:diskcondleadingtogiants}), but is not important in determining the final total mass, because in the end, gas makes up for most of the mass of a giant planet, and not solids. Therefore, a high metallicity mainly allows a larger number of high mass planets, but not of a higher mass (except for very large masses, see sect. \ref{maximalmassesandmetallicity}). This increase in frequency is of course the underlying reason for the observed metallicity effect, i.e. the increase of the detection rate of giant planets with [Fe/H] (see Sec. \ref{sect:detectprobmetallicity}). 

\begin{figure*}
      \centering
      \begin{minipage}[lt]{12.1cm}
        \includegraphics[width=\textwidth]{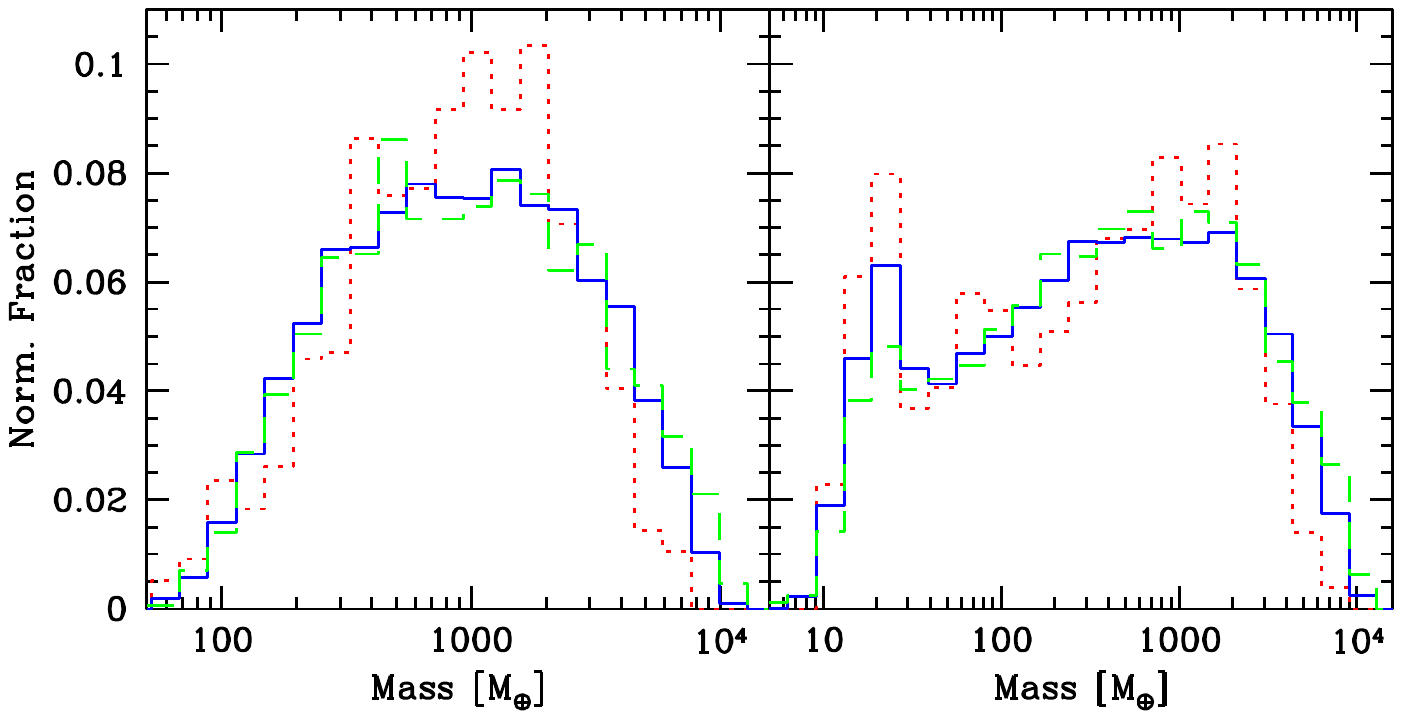}
        \end{minipage}
        \hfill
\begin{minipage}{5.9cm}
      \caption{Planetary mass function binned according to [Fe/H] as in Fig. \ref{fig:imfall}, left panel, but now only for the sub-population of potentially detectable synthetic planets with an RV precision of 10 m/s (left panel) and 1 m/s (right panel). Green dashed lines again represent high, blue solid lines medium and red dotted lines low metallicities, where the threshold metallicities are the same as in Fig. \ref{fig:imfall}.}
       \label{fig:imfobs}
       \end{minipage}

\end{figure*}

Moving down in mass, we see that in the Neptunian mass domain, the dependence of planet frequency on metallicity is weak. This means that no metallicity effect is predicted in this domain.  This is in good agreement with recent observations (Mayor et al. \cite{mayormarmier2011}). Moving further down in mass to the proto-terrestrial domain, the lines are inverted relative to the giant planet domain, which means that an inverse metallicity effect occurs (low mass planets are more frequent at low [Fe/H] compared to high [Fe/H]). Concerning this last point, we must take into account that  in our model, only one embryo can form per disk. The situation that in high [Fe/H] disks, both a giant planet and a ``byproduct'' terrestrial planet form is therefore not possible. This could artificially strengthen the [Fe/H]-low mass planet anti-correlation, so that more secure predictions will be possible with models dropping the one-embryo-per-disk simplification (Alibert et al. in prep.). Observationally, future high precision observations (e.g. with ESPRESSO) will test this prediction. 

One further notes that for all three bins, the highest peak in the mass function is found for the  proto-terrestrial planets, with the clearest dominance at low metallicity. For high metallicity, the second highest peak occurs for the giant planets, whereas for intermediate and even more clearly for low metallicity, the Neptunian planets are responsible for the second highest peak.

\subsubsection{PIMF as function of disk gas mass}\label{sect:pimfdiskgasmass}
The central panel shows that the disk mass distribution directly affects the  shape of the PIMF and not only the height of the peaks of the distribution as metallicity does. For giant planets, a high $\mm$ shifts the formation of giant planets from lower mass to higher masses (compare the blue and green line). For lower disk masses, massive giant planets  ($\gtrsim 6\, \mj$) cannot form at all. In this case, there is simply not enough gaseous material available. The distribution of intermediate-mass  planets ($30\,\mearth \lesssim M\lesssim1\,\mj$), on the other hand, is little affected. We conclude that in our model there is a direct correlation between the (maximal) mass of giant planets and the disk gas masses. This is essentially due to the fact that  giant planets accrete most of their mass in a regime where their accretion rate is proportional to the disk mass, see  the  discussion in Section \ref{sect:maximalmassesanddiskmass}. 

In Paper III, we showed that, in our alpha-disk model, the disk mass has to be scaled roughly linearly with $\mstar$ in order to reproduce the observed correlation between the star's accretion rate and its mass. As shown in this paper, this translated into the formation of planets of a higher mass orbiting stars of a larger mass, in good agreement with observation (Lovis \& Mayor \cite{lovismayor2007}). Even though we did not vary the stellar mass in the present study, the fact that more massive planets  form in disks taken from the high-mass end of the distribution stems from the same reason. Unfortunately, in practice we cannot infer the primordial disk mass for any actually detected exoplanet and so this correlation is difficult to test observational in contrast to the metallicity correlation. The panel further shows that the relative importance of Neptune-like planets strongly depends on the disk gas mass.  Neptune-like planets form particularly easily in disks with small primordial disk masses while they do not seem to be able to form in massive disks. This again is a consequence of the same effect discussed above and already pointed out in Paper III.

\subsubsection{PIMF as function of disk lifetime}
The third panel at the right of Fig. \ref{fig:imfall} finally shows the disk lifetime. The disk lifetime has a twofold influence. In disks with longer lifetimes, cores will be able to grow to the critical mass and accrete gas in a runaway fashion even for relatively low solid surface density. Therefore, the disk lifetime acts as a threshold for giant planet formation, similar to the metallicity. The lifetime of  disks however also affects the total mass of the planets similar to the disk mass, as, to first order, the planet's final mass will be equal to the accretion rate times the duration of the accretion phase. This twofold effect is clearly  seen  in the figure. Giant planets formed in long-lived disks are both numerous and of a higher mass. Compared to metallicity and disk masses, the disk lifetimes has thus a more complex influence on the resulting planet population. The disk lifetime both scales and distorts the shape of the PIMF. Similarly to the primordial disk mass, it is difficult to deduce the disk lifetime for any give observed system. We come back to that in sect. \ref{sect:detectprob}.  

\subsection{Observable distribution as function of [Fe/H]}\label{sect:imfobservabledist}
It is interesting to look whether some of the correlations between disk properties and the underlying mass function (of all planets) discussed in the section above can already be seen in the observational data we have today, which represent only a small fraction of all existing planets.  For this, we plot in Fig. \ref{fig:imfobs} the mass histogram binned according to [Fe/H], but now including only planets detectable by a 10-year duration radial velocity (RV) survey with a precision of either 10 m/s (left panel) or 1 m/s precision (right panel). Note that, like as in Paper III, we use a very simple velocity amplitude cut-off criterium to determine detectability.    

\subsubsection{10 m/s radial velocity precision}
The figure shows that at a precision of 10 m/s, the mass distribution has a very similar shape for all three metallicity bins, as expected from the discussion above and the previous work by Ida \& Lin (\cite{idalin2004b}). However, a closer look indicates that the low [Fe/H] bin has a somewhat narrower distribution, and that there is a certain systematic difference at the upper mass end (see the next section). However,  the medium and high metallicity bins to which most of the present day known planet population belongs, are very similar. It is therefore not surprising that, given the relatively small number of observed planets compared to the large number of synthetic planets used here, no significant correlation between the shape of the mass distribution of giant planets and the host star metallicity has been noticed thus far. 

\subsubsection{1 m/s radial velocity precision}
At a precision of 1 m/s on the contrary, the influence of [Fe/H] on the PIMF becomes visible as measured by the relative frequency of Neptunian versus Jovian planets, a trend that has been observed (Udry et al. \cite{udryetal2006}). We find that the ratio of the number of Neptune-mass to Jupiter-mass planets strongly correlates with the metallicity. For the low metallicity, the Neptune-mass planets are almost as numerous as Jupiter-mass planets while for high metallicity disks they are significantly less frequent. This corresponds well with the observed result (Sousa et al. \cite{sousaetal2008}).  The numerical values of the ratios of synthetic Jupiter-to-Neptune number of planets (dividing mass defined at 30 $\mearth$) obtained for a 10 year search at 1 m/s is in our nominal simulation 4.7, 5.8 and 7.6 for the low, medium and high [Fe/H] bin respectively. These values are difficult directly with Sousa et al. (\cite{sousaetal2008}) due to several reasons (differing [Fe/H] bins, differing primary masses, differing survey duration). However, these authors find values for the Jupiter-to-Neptune ratio of 1.6-5 for [Fe/H]$<-0.15$, and 7.5-30 for [Fe/H]$>0.15$.  Note that the ratios even at 1 m/s precision are higher than those in the full underlying distribution without detection bias.  The reason is that even at 1 m/s many Neptune-like planets remain undetected. Finally, we point out that due to our assumption of a single planet-per-disk simplification, it is very likely that our ratio is an upper limit. A hint in the same direction can also be inferred when comparing our synthetic result with the observed one shown in Fig. 10 of Mayor et al. (\cite{mayormarmier2011}). Otherwise, observed and synthetic result are similar.

From the two plots we also understand that the decrease of the observed mass distribution at 10 m/s towards small masses, while being clearly the consequence of the observational detection bias against low mass planets, is amplified by the actual true decrease of the underlying mass distribution. This degree of depletion of intermediate mass planets ($30-100\, \mearth$) is potentially an important constraint on the gas accretion rate at the beginning of runaway gas accretion. In Mordasini et al. (\cite{mordasinietal2010}), we presented a detailed discussion about various assumptions about gas accretion that directly determine this portion of the theoretical PIMF.

\subsection{(Maximal) planetary masses and metallicity}\label{maximalmassesandmetallicity}
\begin{figure}
\includegraphics[angle=0,width=\columnwidth,origin=br]{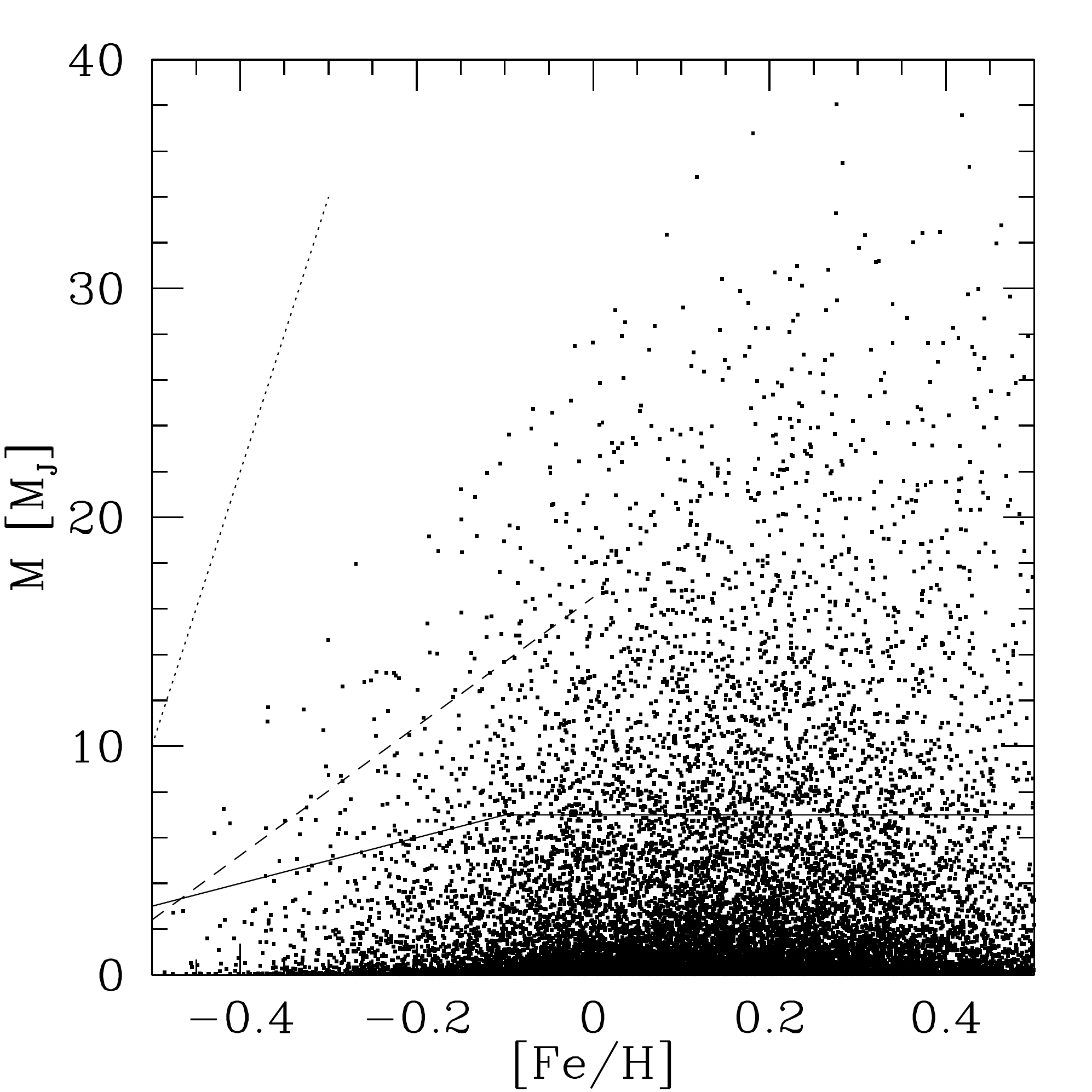}
\caption{The mass of synthetic planets (in Jupiter masses $M_{J}$) as a function of [Fe/H] for the nominal population (small dots). There is an absence of very massive planets around low metallicity stars. In the plot, lines indicate approximatively the region of high mass and low metallicity where no synthetic giant planets are found in other non-nominal populations: with an ice line fixed to 2.7 AU (dotted line), or with the effect of gap formation on the gas accretion rate modeled as in Veras \& Armitage (\cite{verasarmitage2004}) (solid line). The dashed line shows the limit given in Fischer \& Valenti (\cite{fischervalenti2005}).}\label{fig:mmax}
\end{figure}

To emphasize the importance of the metallicity (measured in our model by the ratio of dust-to-gas in the disk) on the resulting high-end of planetary masses, we show in Figure \ref{fig:mmax} the mass of synthetic planets (in units of Jupiter masses) as a function of metallicity. While we have indicated in Sect. \ref{subsubsect:imf} that metallicity does not change significantly the distribution of the mass for the bulk of the population, we see here that the metallicity determines the maximum mass a planet\footnote{We here call for simplicity all objects that form in the population synthesis planets, even if their mass is larger than the deuterium burning  limit at about 13 $\mj.$} can grow to in a given disk, in particular for subsolar metallicities.   This can be understood as follows. To grow to a very high planetary mass, a critical core must form before viscous evolution and photoevaporation of the disk have had time to significantly deplete the disk mass. Such a very early start is not possible in a low metallicity environment in which growing large cores takes longer. 

This second order correlation via the core formation timescale has been checked and confirmed by  artificially reducing the starting time of the seed embryos by a factor of two. In this test,  the maximal mass of planets at [Fe/H]=-0.4 increases to about 18 $\mj$ compared to $\sim7 \mj$ in the nominal case. 

\subsubsection{Effect caused by the iceline position}\label{sect:effectcausedbyaice}
A less pronounced absence of massive planets at low metallicity can also be obtained if we fix the position of the ice line at 2.7 AU independently of $\mm$, as expected for an optically thin disk irradiated from a 1 $\msun$ star (Ida \& Lin \cite{idalin2004a}). This difference is due to an interesting chain of correlations: \\
1)We recall that in the nominal case, the position of the ice line is an increasing function of disk mass.\\
2)At low metallicities, high disk gas masses are needed for giant planet formation (Matsuo et al. \cite{matsuoetal2007}; Sect. \ref{sect:compensationeffects}). Thus, the ice line position in low metallicity disks forming giant planets will typically be located at large distances. \\
3)Hence, the typical starting position of giant planets-to-be will also be large at low [Fe/H] ($\astart\gtrsim\aice$, see Sect. \ref{sect:startingposition}). \\
4)This distant starting position implies a  slower growth of the core (e.g. Paper I), disadvantageous for successful giant planet formation.\\
Thus, a positive correlation of the disk gas mass and the iceline position renders giant planet formation at low [Fe/H] even more difficult, and makes it impossible to populate the upper left corner of Fig. \ref{fig:mmax} with the nominal model. It also means that there is a difference between a metal poor-gas rich and a metal rich-gas poor environment, even if the solid surface density beyond the ice line are in principle the same. A low [Fe/H] cannot in all circumstances be compensated a high $\mm$. 

In the case with an ice line independent of $\mm$, this complex chain of correlations between [Fe/H], $\mm$, $\aice$, $\astart$ and $\tstart$ is broken, and planets with about 10 $\mj$ can still form at [Fe/H]=-0.5, and that maximal masses become independent of [Fe/H] already for [Fe/H]$\gtrsim-0.3$. The upper limit of possible masses is indicated for this case by the dotted line in Figure \ref{fig:mmax}. It is clear that this chain of correlations depends quite sensitively on a number of specific model assumptions, making it a less robust prediction.

\subsubsection{Very massive planets}\label{sect:verymassiveplanets}
The absence of very massive companions formed by core accretion at very low metallicities is also seen in the models of Ida \& Lin (\cite{idalin2004b}) and Matsuo et al. (\cite{matsuoetal2007}). However, compared to these works, planets of significantly higher mass can form in our simulations. This is a direct consequence of the fact that we do not limit gas accretion due to gap formation, as mentioned in Sect. \ref{sect:relevantmodelassumptions}. If  we instead  limit  in a non-nominal population the gas accretion rate by giant planets due to gap formation by using the fit of Veras \& Armitage (\cite{verasarmitage2004}), the maximal mass of all planets in the population  are reduced to $\sim7\mj$. The absence of the most massive objects at low [Fe/H] would remain similar but with weaker dependence of the maximal mass on [Fe/H] when scaled to smaller absolute masses. The approximative limiting envelope for such a population is also shown in Fig. \ref{fig:mmax} as a solid line.  This envelope is now similar to the one of  Matsuo et al. (\cite{matsuoetal2007}). We thus see that while the tendency towards an absence of very massive planets at low [Fe/H] is a general prediction of the core accretion theory, the quantitative results depend on specific model settings. 

When studying the figure one should keep in mind that very massive planets ($\gtrsim10\mj$) are in fact very rare outcomes in our simulations, in agreement with the observed ``brown dwarf desert'' as discussed in Paper II. Figure \ref{fig:mmax} may provide the somewhat misleading impression that these are common objects while in fact these objects appear only because the underlying synthetic population is extremely large (about 200\ 000 initial conditions). Even if such a population vastly exceeds the actually observed population, these high numbers of planets are required to investigate the different correlations some depending on two variables (like disk mass and [Fe/H]).

\subsubsection{Comparison with observation}\label{sect:compobsmmaxfeh}
The absence of very high mass planets (or of a very high total mass locked up in planets in multi-planet systems) at low metallicities was noted also observationally in Udry et al. (\cite{udryetal2002}), Santos et al. (\cite{santosetal2003}) and Fischer \& Valenti (\cite{fischervalenti2005}). The dashed line in Fig. \ref{fig:mmax} is taken from the latter work. It is clear that such an absence could also simply be a small number effect: Giant planets around low [Fe/H] hosts are rare, and high mass giant planets ($\gtrsim5 \mj$) are rare at all metallicities, which could combine into the observed paucity. It is interesting to compare our results with the observational database especially since the latter has been significantly extended since the above mentioned papers have been published. 

\begin{figure}
 \includegraphics[angle=0,width=\columnwidth,origin=br]{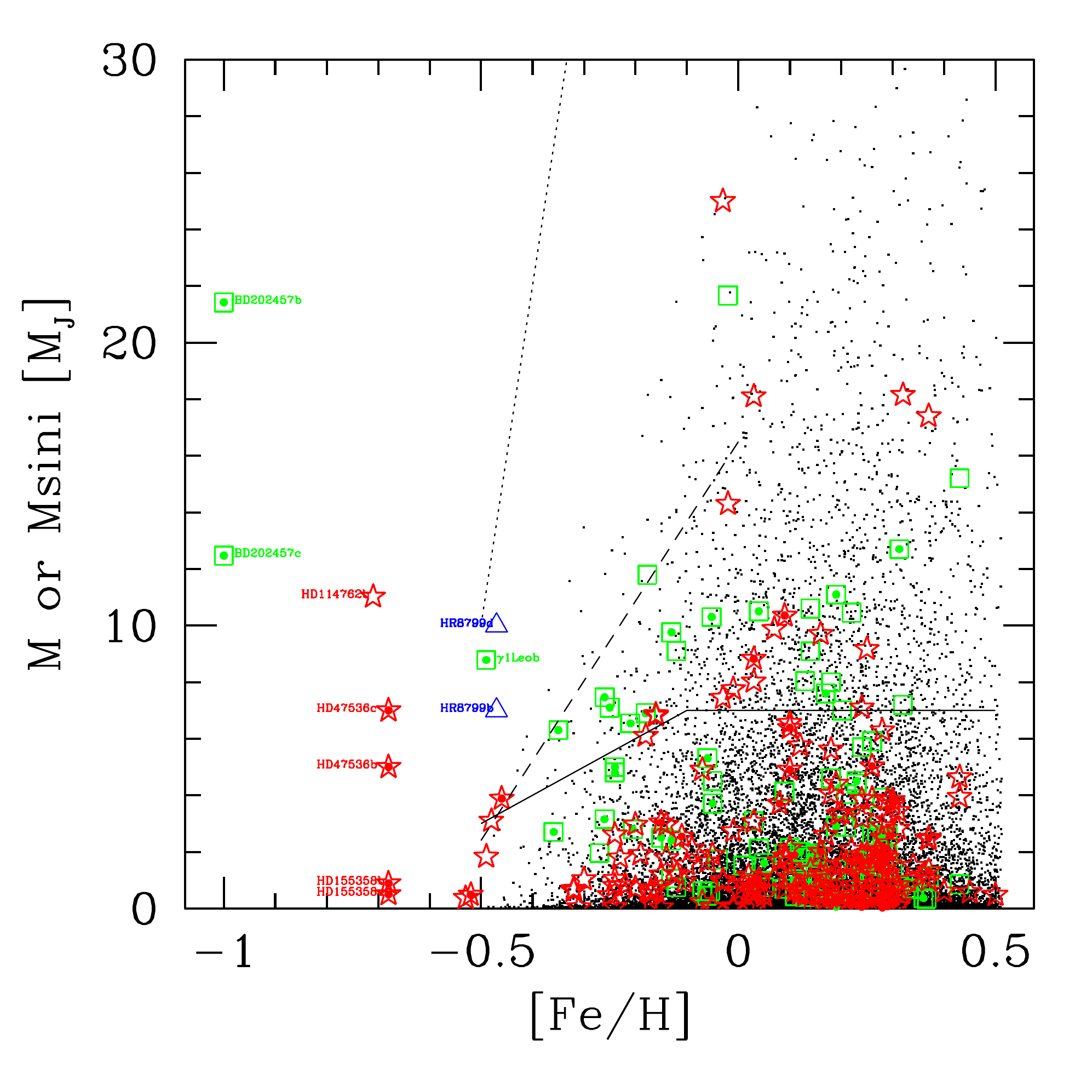}
 \caption{Observed and synthetic planetary masses as a function of metallicity. Black dots are synthetic planets in the nominal model, while  solid, dashed and dotted lines again indicate the limiting envelope of non-nominal models, as in Fig. \ref{fig:mmax}.  Red stars are observed RV companions around stars with $ 0.8<\mstar/\msun<1.2$. Green squares are observed RV planets around stars with a mass larger than 1.2 $\msun$. For both cases, a point in the middle of the symbol indicates that the star is a subgiant or giant. Blue triangles show planets around HR 8799. Names of relevant objects are given in the plot.}\label{fig:mfehobs}
\end{figure}

Such a comparison is presented in Fig. \ref{fig:mfehobs} which shows the masses of actual and synthetic planets as a function of metallicity. Observational data was taken from J. Schneider’s Extrasolar Planet Encyclopedia. Note that this database does not include companions larger than about 20 $\mj$. However, such companions are anyway extremely rare, both in the model (only 0.1\% of the synthetic planets have a mass larger than 20 $\mj$) but also observationally. As a result, our observational knowledge of this mass range, where several formation mechanism could contribute, still suffers from low number statistics (see also Sozzetti \& Desidera \cite{sozzettidesidera2010}).

In Fig. \ref{fig:mfehobs}, red stars symbolize planets detected by the radial velocity method around primaries with a mass between 0.8 and 1.2 $\msun$, which are the most relevant cases (in the model $\mstar=1\,\msun$). Green squares are companions to stars more massive than this, also detected by radial velocity. Stars which have evolved off the main sequence (if known) are marked in the figure with a dot. Note that stellar evolution could be of relevance here, as it might invalidate our underlying assumption that the photospheric composition measured today correlates with the bulk composition of the disk material at formation. This assumption could be invalid in case of enhanced heavy element settling, which for the sun seems  to have already lead to a reduction of the present day Z as compared to the primordial Z$_{0}$ by more than 10\%  (Lodders \cite{lodders2003}). 

Looking at  Fig. \ref{fig:mfehobs}, we see that the bulk of all observationally detected planets falls into regions of the plot where synthetic planets are also found. Hence, from this observation we could conclude that core accretion can account for almost all planets currently known. This conclusion was already reached by  Matsuo et al. (\cite{matsuoetal2007}). It should be noted that surveys that look at very metal poor stars do exist (Santos et al. \cite{santosetal2009}). 

\subsubsection{Relevant individual objects}
However, eight RV planets orbiting five stars clearly lie outside the region where giant planets form orbiting a 1 $\msun$ star in our nominal model. This group is characterized by a much higher fraction of giant stars and of stars more massive than 1.2 $\mstar$ than the full sample.  It also has a high multiplicity for giant planets. As some of the objects are quite peculiar, it is worth to discuss individually some of these objects. 

\object{HD\,155358} has two planetary companions of rather low mass ($\lesssim 1\,\mj$) at 0.6 and 1.2 AU (Cochran et al. \cite{cochranetal2007}). The star has a mass of $0.87\pm0.07\msun$ and a [Fe/H]$=-0.68\pm0.07$. \object{HD\,155358} is a particular star (Fuhrmann \& Bernkopf \cite{fuhrmannbernkopf2008}). It is a very old, thick disk subgiant with a chemical composition very far from scaled solar composition. It is significantly enriched in alpha-chain nuclei. [Mg/H] is for example with -0.36 much higher than [Fe/H]. What matters for the ability to form a sufficiently massive core is the surface density of all condensible elements beyond the ice line which can attribute important quantities of matter, in particular the $\alpha$ elements O, Si and Mg (Dodson-Robinson et al. \cite{dodsonrobinsonetal2006}; Gonzalez \cite{gonzalez2009}). Due to the enrichment in $\alpha$ elements, the disk around \object{HD\,155358} was depleted in planetesimals maybe only by a factor 2 below a [Fe/H]=0 solar composition disk, equivalent to a [Fe/H]$\sim-0.3$ if it were to have a scaled solar composition.  At such a value, forming the anyway low mass giant planets is certainly possible with the core accretion mechanism. 

\object{HD\,114762} with its $\msini= 11\,\mj$, a=0.36 AU companion  (Latham et al. \cite{lathametal1989}) was discussed in this context already by Udry et al. (\cite{udryetal2002}). It has been often considered that the system is seen nearly pole on, so that  the companion might in fact be a late M dwarf (Cochran et al. \cite{cochranetal1991}; Hale \cite{hale1995}). An alternative hypothesis is based on the finding that \object{HD\,114762} is chemically and evolutionary very similar to \object{HD\,155358} (Fuhrmann \& Bernkopf \cite{fuhrmannbernkopf2008}) which would mean that its effective surface density of planetesimals was much higher than one would infer from the low [Fe/H] and scaled solar composition. 

\object{HD\,47536} is an old K1III, [Fe/H]=-0.68 giant which is orbited by one (Setiawan et al. \cite{setiawanetal2003}), possibly two (Setiawan et al. \cite{setiawanetal2008}) quite massive companions (5 and 7 $\mj$) inside a few AU. While  Setiawan et al. (\cite{setiawanetal2003}) originally quoted a possible stellar mass of  about 1 to 3 $\msun$, da Silva et al. (\cite{dasilvaetal2006}) more recently determined a mass of $0.94\pm0.08\mstar$. With this primary mass, the object falls clearly out of the envelope of synthetic planets in our model. In particular if its two-planet configuration is confirmed, this is  an interesting object to study with formation models, as one then cannot invoke, due to orbital stability,  a nearly pole on orientation as a possible explanation.  Also for the direct collapse model this case  is probably not obvious to explain due to the small semimajor axis of the planets (about 1.5 AU for the inner planet), see e.g. Boley (\cite{boley2009}). 

\object{BD+20\,2457} is a K2II giant, with an estimated mass of $2.8\pm1.5\msun$, a very low metallicity of [Fe/H]=$-1.00\pm0.07$ with two very massive companions in tight orbits (Niedzielski et al. \cite{niedzielskietal2009}). Even if the very different primary mass makes a comparison with the results here difficult,  these companions seem to be far from the possible parameter space for core accretion. 

The object around \object{$\gamma^{1}$ Leo A} (\object{HIP  50583}) was found to have a projected mass of about 8.8 $\mj$ (Han et al. \cite{hanetal2010}). The orbit of the companion was however recently  astrometrically detected  by Reffert \& Quirrenbach (\cite{reffert2011}) in the \textit{Hipparcos} data. Their results indicate a clearly larger actual mass of about 66 $\mj$ (with substantial incertitude). This moves the companion out of the relevant mass domain.

The plot also includes \object{HR 8799} with a measured [Fe/H]=-0.47 (which might however not reflect the initial metallicity, Marois et al. \cite{maroisetal2008}) and planets detected by direct imaging i.e. at large orbital distances (Marois et al. \cite{maroisetal2008, maroisetal2010}). HR 8799 also probably has a mass larger than 1.2 $\msun$ ($1.5\pm0.3\msun$), so the synthetic population does not apply directly, but it is nevertheless interesting to note that these planets  come to lie in a region in the [Fe/H]-mass plane which is at or close to the limit where giant planets can originate from core accretion. This is a finding  independent from the fact that the large semimajor axes of these planets makes core accretion as the formation mechanism difficult  (Dodson-Robinson et al. \cite{dodsonrobinsonetal2009}), if no additional mechanism causing outward displacement like scattering is acting.  

\subsubsection{Summary concerning the [Fe/H]-$M$ correlation}
In summary we see that there are only extremely few examples of bona fide, massive companions ($5\lesssim M \lesssim 20\,\mj$) orbiting solar-like, main sequence stars at small orbital distances which do not fall into the [Fe/H]-$M$ parameter space covered by our implementation of the core accretion model. 
 
The solid line in Fig. \ref{fig:mfehobs} shows also the limiting envelope of synthetic planetary masses derived in the case where the gas accretion rate due to gap formation is limited by using the simple one parameter fit of  Veras \& Armitage (\cite{verasarmitage2004}) to hydrodynamical simulations of Lubow et al. (\cite{lubowetal1999}). It is clear that such a simple fit can only be a rough approximation of the real effect and that e.g. disk viscosity also has an influence on the degree of quenching of gas accretion (Lissauer et al. \cite{lissaueretal2009}). But the significant number of planets lying above the solid line indicates that mechanisms like the eccentric instability (Kley \& Dirksen \cite{kleydirksen2006}) that allow growth beyond the gap barrier seem to play an important role in nature. 

We finally note that the most massive synthetic planets could also, at least partially, be an artifact of the one embryo per disk approximation. In a disk where several protoplanets concurrently grow, planets compete for gas and ejection can even remove some planets (Thommes et al. \cite{thommesetal2008}). In this sense, looking at the total mass in the system should be a more adequate quantity than individual planet masses.

\subsection{Planetary masses and disk mass}\label{sect:maximalmassesanddiskmass}
\begin{figure}
\includegraphics[angle=0,width=\columnwidth,origin=br]{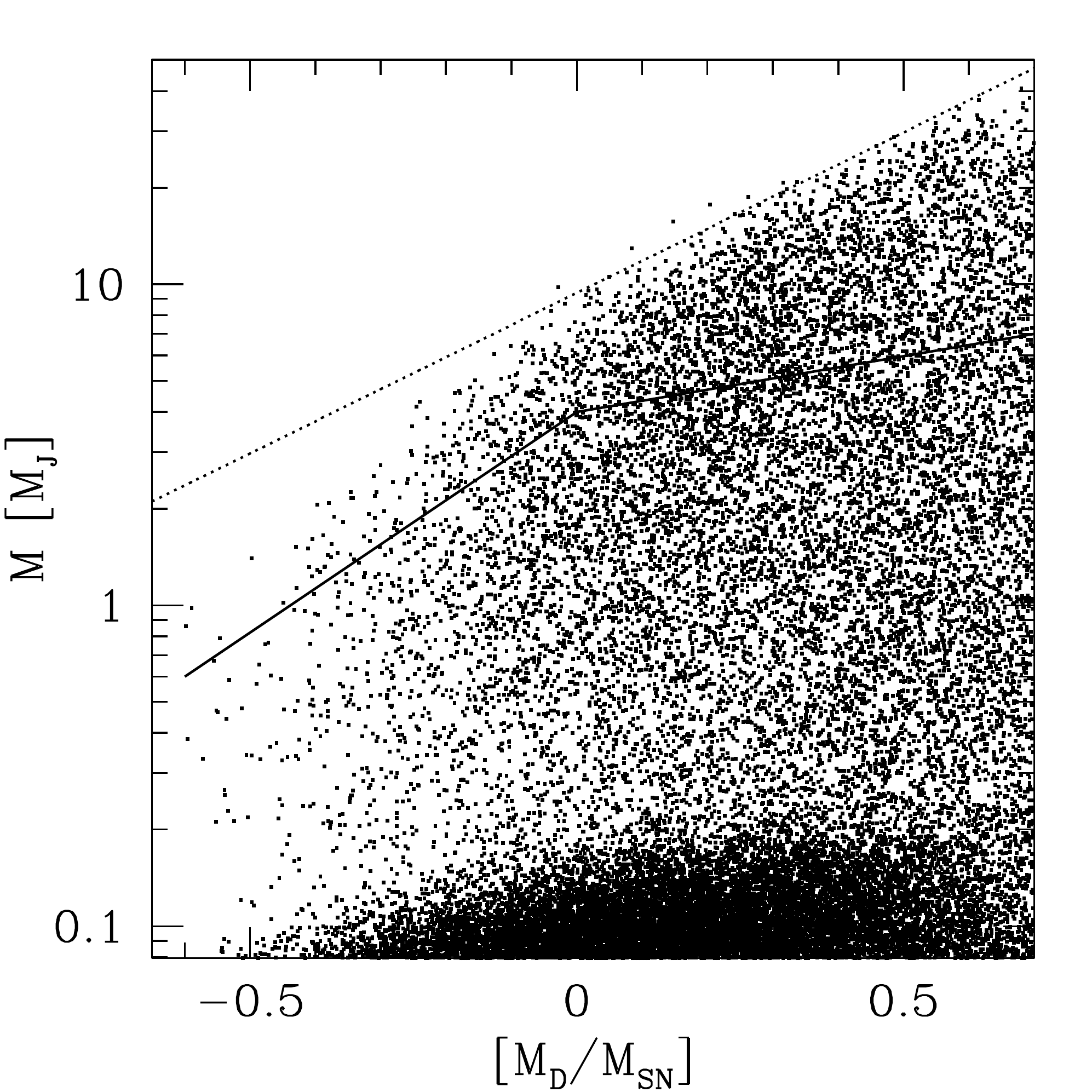}
\caption{The mass of synthetic planets (in Jupiter masses) as a function of the initial disk mass for the nominal population (black dots). The dotted line indicates a mass half as large as the initial gas disk mass, showing the linear correlation of disk mass and maximal planet mass for most of the domain. The solid line shows the upper limiting envelope in a non-nominal population where the  limiting effect of gap formation on the gas accretion rate is taken into account, using the fit of Veras \& Armitage (\cite{verasarmitage2004}).}\label{fig:mmaxSigma}
\end{figure}

Figure \ref{fig:mmaxSigma} shows planetary masses as a function of the initial disk gas mass, again in units of Jupiter masses to focus on the massive planets. Here we find a different behavior than for [Fe/H]: As expected from the discussion of the influence of $\mm$ on the PIMF, we see that there is a positive correlation of disk mass and planetary mass over the full domain of disk masses considered here. This is illustrated by the dotted line corresponding to a mass that is equal half the initial disk mass. This shows that there is, for the domain of $\mm\gtrsim-0.2$ an approximatively linear correlation between the maximal planet mass and the disk mass.  The reason for this is that giant planets accrete in our model (see Mordasini et al. \cite{mordasinietal2009a}) most of their mass in the disk limited accretion regime where the planetary gas accretion rate is assumed to be the same as in the disk ($\dot{M}_{\rm env}=\dot{M}_{\rm disk}=3 \pi \nu \Sigma$), which is itself proportional to the initial disk mass for a given distance and moment in time, at least if the disk evolution is close to self-similarity solutions (Hartmann et al. \cite{hartmannetal1998}). 

The different influence of [Fe/H] and disk gas masses on the mass of giant planets can be quantified: For planets with a mass larger than 100 $\mearth$, one finds that the median planetary mass grows for an increase of [Fe/H] from -0.45 to 0.45 only by about a factor 1.5 (from 512 to 747 $\mearth$). For an increase over the same $\mm$ domain in contrast, the median giant planet mass increases from 182 to 814 $\mearth$ i.e. by about a factor 4.5.

For $\mm\lesssim-0.2$, the maximal masses  decrease faster then linearly with decreasing $\mm$ and are thus smaller than indicated by the dotted line. This is because at such low disk masses, the  low resulting solid surface $\sigmas=\fpg\sigmanorm$ densities becomes important in a second order effect, so that the final planetary mass is no more just determined by the amount of gas that can be accreted, but also by the time needed to form a supercritical core. This is an effect that is analogous to the dependence of the maximal mass seen at low [Fe/H].  Note that the exact value of the highest efficiency of converting disk gas into planetary envelope material (here about 0.5) depends on the exact treatment of the back-reaction of the planet's accretion on the disk, but is found to lie in a domain of 0.3 to 0.5. The linear correlation between disk mass and maximal planetary mass is however not affected by different treatments. This indicates that a positive correlation of disk mass and giant planet mass is a stable prediction of the core accretion theory, while the quantitative degree is model dependent. For the large majority of giant planets ($M\geq1\,\mj$) the efficiency of converting disk gas into planetary material is lower and to order of magnitude 0.1, but with a very wide spread of possible values, depending on e.g. the photoevaporation rate, the metallicity or the disk gas mass itself.  
   
The solid line in Figure \ref{fig:mmaxSigma} indicates again the approximative upper envelope obtained for the non-nominal population using gas accretion rates limited because of gap formation according to the Veras \& Armitage (\cite{verasarmitage2004}) fit. In this case, the planetary gas accretion rate is still proportional to  $\dot{M}_{\rm disk}$ but decreases exponentially as $\exp(-\mplanet/1.5\,\mj)$ with planet mass until a floor value of $\dot{M}_{\rm env}=0.04\dot{M}_{\rm disk}$ is hit. The consequence is as shown by the figure that the clear correlation of disk mass and planet mass is now broken, as expected from the functional form of the equation: One finds that the maximal mass now depends only weakly on the disk mass, approximatively as $(M_{\rm D}/M_{\rm SN})^{1/4}$ for $M_{\rm D}\gtrsim M_{\rm SN}$. A weak correlation remains because of the floor accretion rate. At lower disk masses, the correlation is somewhat stronger, for the same reason as for the nominal population.      

\begin{figure*}
\begin{minipage}[t]{0.5\textwidth}
\includegraphics[width=\textwidth]{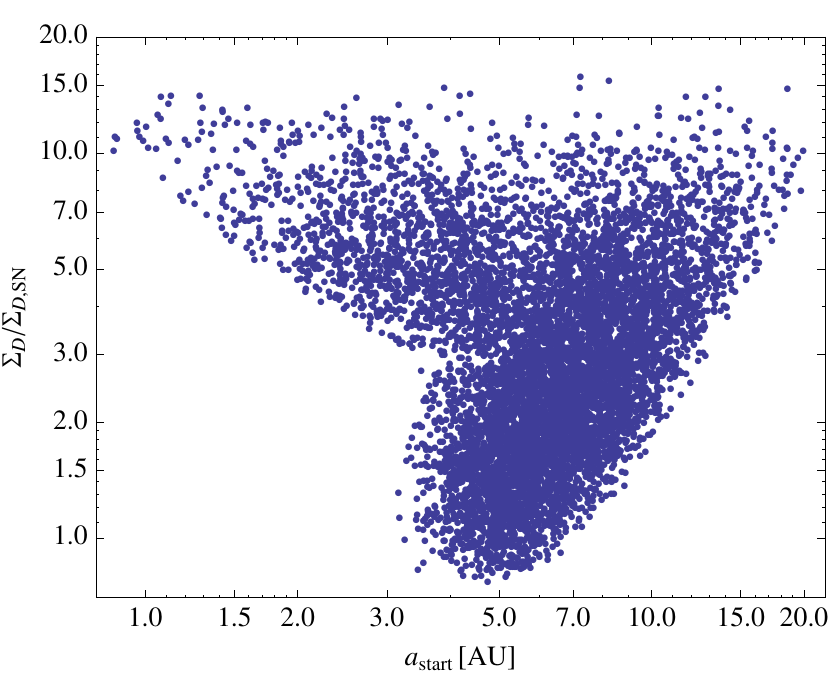}
\end{minipage}
\hfill
\begin{minipage}[t]{0.5\textwidth}
\includegraphics[width=\textwidth]{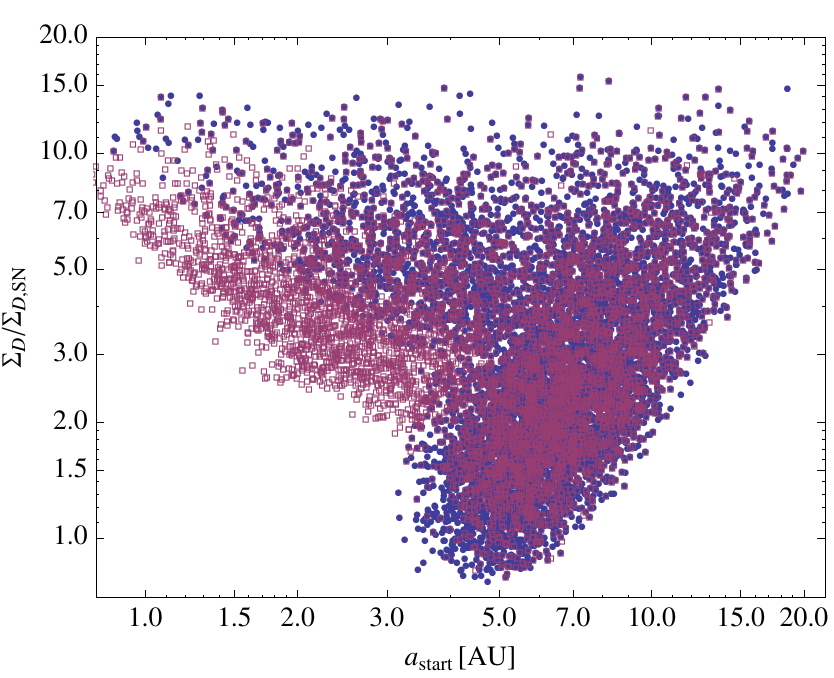}
\end{minipage}
\begin{minipage}[t]{0.5\textwidth}
\includegraphics[width=\textwidth]{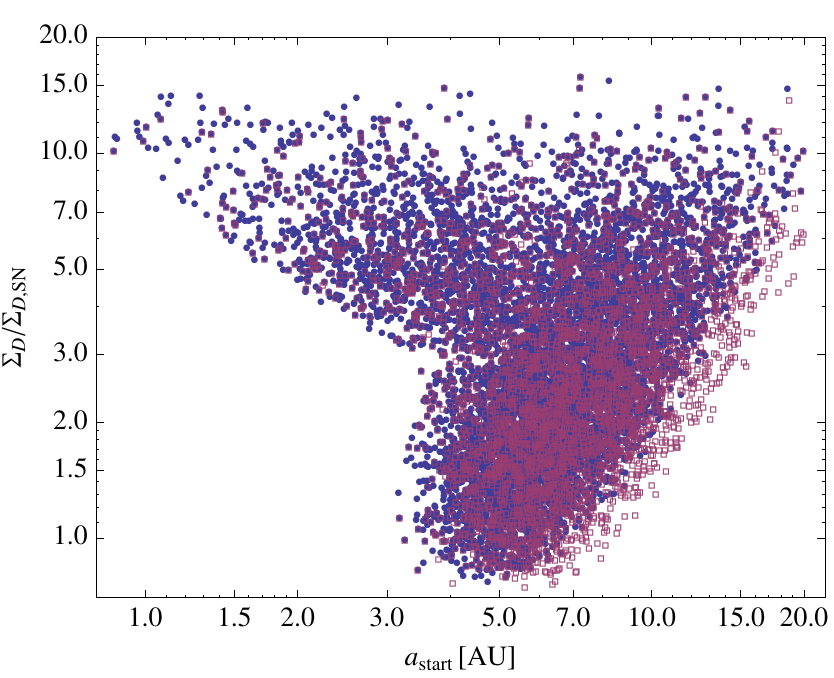}
\end{minipage}
\hfill
\begin{minipage}[t]{0.5\textwidth}
\includegraphics[width=\textwidth]{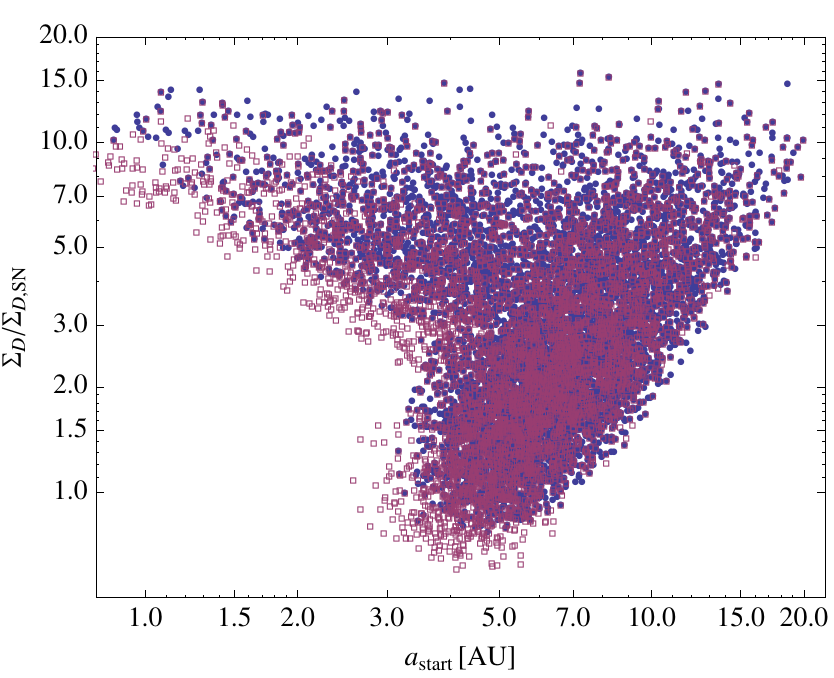}
\end{minipage}
 \caption{Relative planetesimal surface density $\sigmas/\sigmassn=\fpg\sigmanorm/(\fpgsun\Sigma_{\rm 0,SN})$  at 5.2 AU leading to the formation of a planet with a final mass $M\geq300\,\mearth$, as function of the starting position of the embryo $\astart$. A value of   $\sigmas/\sigmassn=1$ corresponds to an initial planetesimal surface density of  8 g/cm$^{2}$ at $\anorm=5.2$ AU, while a value of for example 3 corresponds at e.g. 10 AU to 9 g/cm$^{2}$. The top left panel shows the nominal population only (blue dots). The three other panels show non-nominal population as open red squares which overlay the nominal population that is also shown for comparison, still with blue dots. Top right: Population with faster type I migration ($\f1=0.1$). Bottom left: Population where  $\tstart$ is reduced by 50\%. Bottom right: Population where the opacity in the planetary envelope is 2\% of the interstellar value.}\label{fig:diskcondgiantsastartsd} 
\end{figure*}

The fact that for the nominal population, maximal masses are linearly proportional to disk masses is interesting as one therefore expects to see an imprint of the disk gas mass distribution on the planetary mass distribution. This is reminiscent of the situation for stars,  where the protostellar core mass function CMF has a remarkably similar functional form as the stellar IMF (e.g. McKee \& Ostriker \cite{mckeeostriker2007}). This question will be addressed further in forthcoming work. 

\section{Disk conditions leading to giant planet formation}\label{sect:diskcondleadingtogiants}

\subsection{Minimal solid surface density}\label{sect:minimalssd}

With the population synthesis calculations, one can study a posteriori  which combinations of disk properties allow the formation of giant planets, and in particular which are the most extreme ones. Comparable studies have been done in Kornet et al. (\cite{kornetetal2006}), Dodson-Robinson et al. (\cite{dodsonrobinsonetal2006}), Ikoma et al. (\cite{ikomaetal2000}) or Thommes et al. (\cite{thommesetal2008}). 

Figure \ref{fig:diskcondgiantsastartsd} shows the relative planetesimal surface density  $\sigmas/\sigmassn$ at 5.2 AU, where $\sigmas=\fpg\sigmanorm$ is the planetesimal surface density in a given disk and $\sigmassn=\fpgsun\Sigma_{\rm 0,SN}=8$ g/cm$^{2}$ is the planetesimal surface density in our solar nebula reference disk, which lead to the formation of a giant planet ($M\geq300\,\mearth$), as a function of the starting position of the seed embryo. The core accretion rate $\mdotcore\propto\sigmas$ (e.g. Alibert et al. \cite{alibertetal2005a}), therefore this quantity is within the core accretion formation paradigm a central control parameter. Blue filled circles represents synthetic planets of the nominal population in all four panels.  A value of e.g. $\sigmas/\sigmassn=3$ thus corresponds to a disk with roughly three times more solids than the amount we assume for the solar nebula. In all disks, the solid surface density scales as $a^{-1.5}$ with a jump of a factor 4 at the ice line, therefore  this value corresponds for a starting position of the embryo of e.g. $\astart=10$ AU (which is outside the ice line for all $\sigmanorm$ considered here) to a local planetesimal surface density at $\astart$ of about $3\times 8\times (10/5.2)^{-1.5}=9$ g/cm$^{2}$.  

\subsubsection{Sweet spot for giant planet formation}
The plot shows that the sweet spot for giant planet formation (i.e. the lowest solid surface densities which allow that) occurs at about 5  AU, at a value of about 0.75, corresponding to 6 g/cm$^{2}$. This value is similar to the result of Dodson-Robinson et al. (\cite{dodsonrobinsonetal2006}) who find that a solid surface density of about 6.5 g/cm$^{2}$ (also at 5.2 AU) is needed to bring an embryo to runaway gas accretion in 7 Myr, which corresponds to the longest living disks in our population (Paper I). At a starting position of 1 AU (which is inside the ice line), a $\sigmas/\sigmassn$ of  at least 10 is needed, corresponding to a local planetesimal surface density of about $0.25\times 10 \times 8 \times  (1/5.2)^{-1.5}\approx240$ g/cm$^{2}$, which is in good agreement with Kornet et al. (\cite{kornetetal2006}). We see that both at smaller and at larger distances, more massive disks of planetesimals are needed. This is in agreement with the works mentioned earlier. The abrupt increase by about a factor 4 at about 3-4 AU simply corresponds to the increase necessary to compensate the decrease of the planetesimal surface density inside of the ice line by an identical factor due to the sublimation of ices. The upper limit of  $\sigmas/\sigmassn$ in the plot corresponds to disks where both $\fpg$ and $\sigmanorm$ come from the upper end of their distributions.

\subsubsection{Isolation mass and timescale effect}\label{sect:isomassandtimescale}
The reasons for the remaining increase of the necessary  $\sigmas$ with both increasing and decreasing $\astart$ i.e. the existence of the sweet spot has been discussed in earlier work (e.g. Kornet et al. \cite{kornetetal2006};  Thommes et al. \cite{thommesetal2008}). Therefore we here only briefly illustrate this by the following two non-nominal populations which are also plotted in Fig. \ref{fig:diskcondgiantsastartsd}: 

The red empty squares in the top right panel correspond to a population with a faster type I migration rate ($\f1=0.1$ instead of 0.001).  In this case, the minimal necessary planetesimal disk mass is reduced at small distances. This shows that at small distances (where the core accretion timescales are short) the small isolation masses (and the associated long Kelvin-Helmholtz timescales for gas accretion) limit giant planet growth. Isolation masses decrease with decreasing semimajor axis which must be compensated by an increasing $\sigmas$. Faster type I migration partially neutralizes this need, as it allows growth beyond the isolation mass and up to the critical mass by increasing the embryo's feeding zone (Alibert et al. \cite{alibertetal2004}). 

The empty red squares in the bottom left panel show the result for a population for which we have arbitrarily reduced $\tstart$ by a factor 2.  In this case at large distances a difference is seen, while in the other parts of the plot, the points lie on top of each other. At large distances, high amounts of solids are available in the embryo's feeding zone, but the core growth timescale is very long. To compensate that, i.e. to still build up a critical core during the disk lifetime, $\sigmas$ must also increase towards the exterior. When we (artificially) speed up the accretion rate (or, equivalently, reduce $\tstart$), this requirement is relaxed. 

\subsubsection{Situation of the solar nebula}
The figure also shows that the solar nebula is not far above the overall minimal necessary $\sigmas$ for giant planet formation. This conclusion is also in agreement with Thommes et al. (\cite{thommesetal2008}).  We find that the overall minimal value is about $\sigmas/\sigmassn=0.75$, i.e. $\sigmas\approx 6$ g/cm$^{2}$. This would correspond to an isolation mass at 5.2 AU of about $5.3\, \mearth$.  Studying the total mass of accreted planetesimals (which can be in the core, or dissolved in the envelope) we however find that the smallest amount of heavy elements in planets more massive than $300\,\mearth$ is clearly larger, namely about $23\,\mearth$.  This is a consequence of migration, and of planetesimal accretion also after runaway gas accretion is triggered. Internal structure modeling of observed transiting extrasolar planets by  Miller \& Fortney (\cite{millerfortney2011}) points to a similar minimal heavy element content of giant planets.

\subsubsection{Impact of the grain opacity}
The bottom right panel of Figure \ref{fig:diskcondgiantsastartsd} finally shows a  population calculated with an opacity of 0.02 times the nominal (interstellar) value in the envelope (e.g. Pollack et al. \cite{pollacketal1996}). Here we find that the minimal necessary value is, as expected, lower, namely about 4.7 g/cm$^{2}$. This reduction is rather small compared to the large effect of the opacity in formation calculations without migration (Pollack et al. \cite{pollacketal1996}; Movshovitz et al. \cite{movshovitzetal2010}). The reason for the rather small influence is that thanks to migration (even with the strongly reduced type I migration rate), cores never get completely cut from a supply of fresh planetesimal to accrete. Therefore, the need for a rapidly growing gaseous envelope (where the opacity matters) in order to expand the solid feeding zone is not as important as in the case of a strict in situ formation. Such a growth mode that is limited only by the accretion of solids, and not the Kelvin-Helmholtz timescale of the envelope, is somewhat related to the situation for Saturn discussed in Dodson-Robinson et al. (\cite{dodsonrobinsonetal2008}). In agreement with this work we see that the reduction of the opacity is important for smaller distances $\lesssim7$ AU.

\subsection{Starting position}\label{sect:startingposition}
\begin{figure}
\includegraphics[width=\columnwidth]{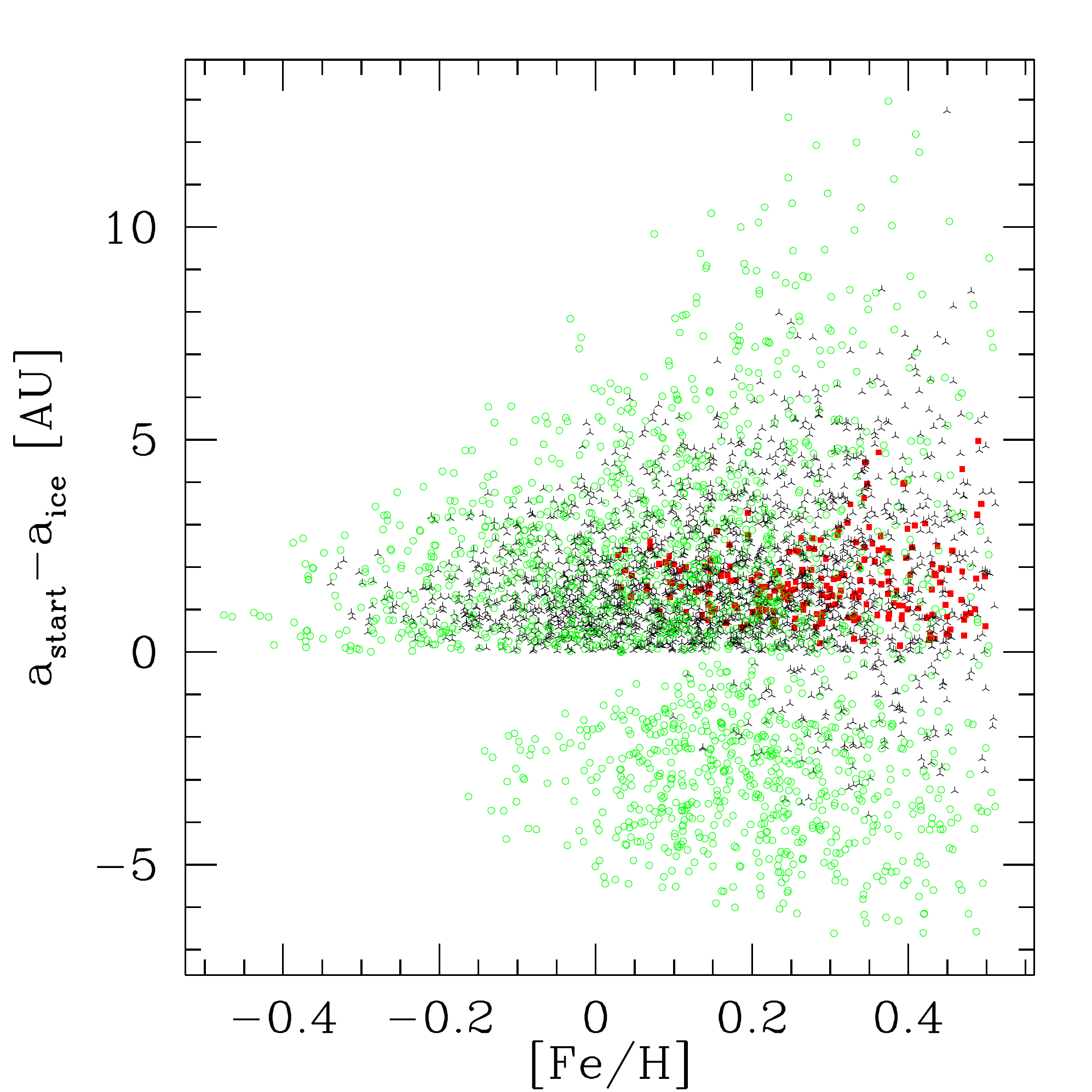}
 \caption{Starting position relative to the ice line of embryos growing eventually to planets with a final mass $M\geq300\,\mearth$ as a function of [Fe/H]. Symbols show the disk gas mass:  Red filled squares are low $\mm<-0.1$. Black triangles are intermediate $\mm$ (-0.1 to 0.4), and green open circles finally are those disk with a large mass $\mm>0.4$.}\label{fig:daiceastart} 
\end{figure}

We have seen in the last section that the location of the ice line is important for the conditions necessary for giant planet formation, and influences in this way the birth place of giant planets. In Fig. \ref{fig:daiceastart} we directly plot the starting position of embryos  $\astart$ which later become planets larger than 300 $\mearth$, relative to the position of the ice line $\aice$ in the corresponding disk, as a function of [Fe/H]. Additionally, the disk mass is color coded. Note that in our simulation, the location of the ice line is independent of [Fe/H] (but dependent on $\sigmanorm$). 

The figure shows that at low metallicities [Fe/H]$\lesssim-0.3$, planets can only start to form in a small region with a width of 2-3 AU just outside the ice line. In this low [Fe/H] disks, only in this region massive cores can form quickly enough. With increasing [Fe/H], the zone expands to larger radii, and for [Fe/H]$\gtrsim-0.15$, giant planets can also form inside the ice line. This corresponds, with the highest $\sigmanorm$ of the distribution (which is necessary, see next section) to a solid surface density about 3.5 times as high as $\sigmassn$ (as also visible in Fig. \ref{fig:diskcondgiantsastartsd}). These results are qualitatively similar to those of Ida \& Lin (\cite{idalin2004a}), except for a somewhat lower numerical value (3.5 instead of 5). The absence of red squares indicating disks with a low gas mass at negative values of $\astart-\aice$ (i.e. inside the ice line) shows that the disk masses must always be rather large for the formation of giant planets inside the ice line, even if [Fe/H] is large. At higher [Fe/H]$\gtrsim0.3$, giant planets can form at almost all semimajor axes  (about 1 AU to 20 AU) if concurrently $\mm$ is high as indicated by the green open circles, but the preferred formation location is still beyond the ice line. Finally, at the highest [Fe/H]$\gtrsim0.4$, the location of the ice line becomes  less important. 

One therefore sees from the figure, that while at low metallicities, pathways to giant planet formation are very restricted and in particular only possible in a small part of the disk, this is not the case for higher metallicities, where giant planets can form all over the disk.

\subsubsection{Anti-correlation of [Fe/H] and $\astart$}\label{sect:anticorrfehastart} 
There is another important chain of correlations regarding the starting positions, which cannot be directly seen in Fig. \ref{fig:daiceastart}, but which was already mentioned in sect. \ref{sect:effectcausedbyaice}: For giant planet formation at low [Fe/H], high $\mm$ are needed for compensation (see below). At high [Fe/H] in contrast, low $\mm$ are sufficient. As the distribution of disk gas masses peaks at values that are close to the lower limit of values allowing giant planet formation,  the typical $\mm$ which leads at a high metallicity to the formation of a giant planet is a rather low $\mm$.  As seen in the figure, also at higher [Fe/H], the ice line remains to be the typical formation location. This means that if low disk masses correspond to low values of $\aice$ (as we assume in the nominal case), there is a negative correlation of [Fe/H] and the typical $\astart$ of giant planets-to-be. So in terms of absolute values of $\astart$ (not relative to the ice line), even if giant planets also come in higher [Fe/H] environments typically from beyond the ice line, their starting positions are smaller in absolute terms, as the typical $\aice$ are smaller. The magnitude of this effect can be quite significant, and corresponds to a typical difference over the 1 dex interval  [Fe/H] covers of about 3 to 4 AU. If $\aice$ is in contrast independent of $\sigmanorm$ (which could be possible due to a dead zone which we do not include here, or if the disk is optically thin), this effect does not exist. It is therefore a finding sensitive to model assumptions.

\subsection{Compensation effects between disk mass and [Fe/H]}\label{sect:compensationeffects}
\begin{figure}
\includegraphics[width=\columnwidth,origin=left]{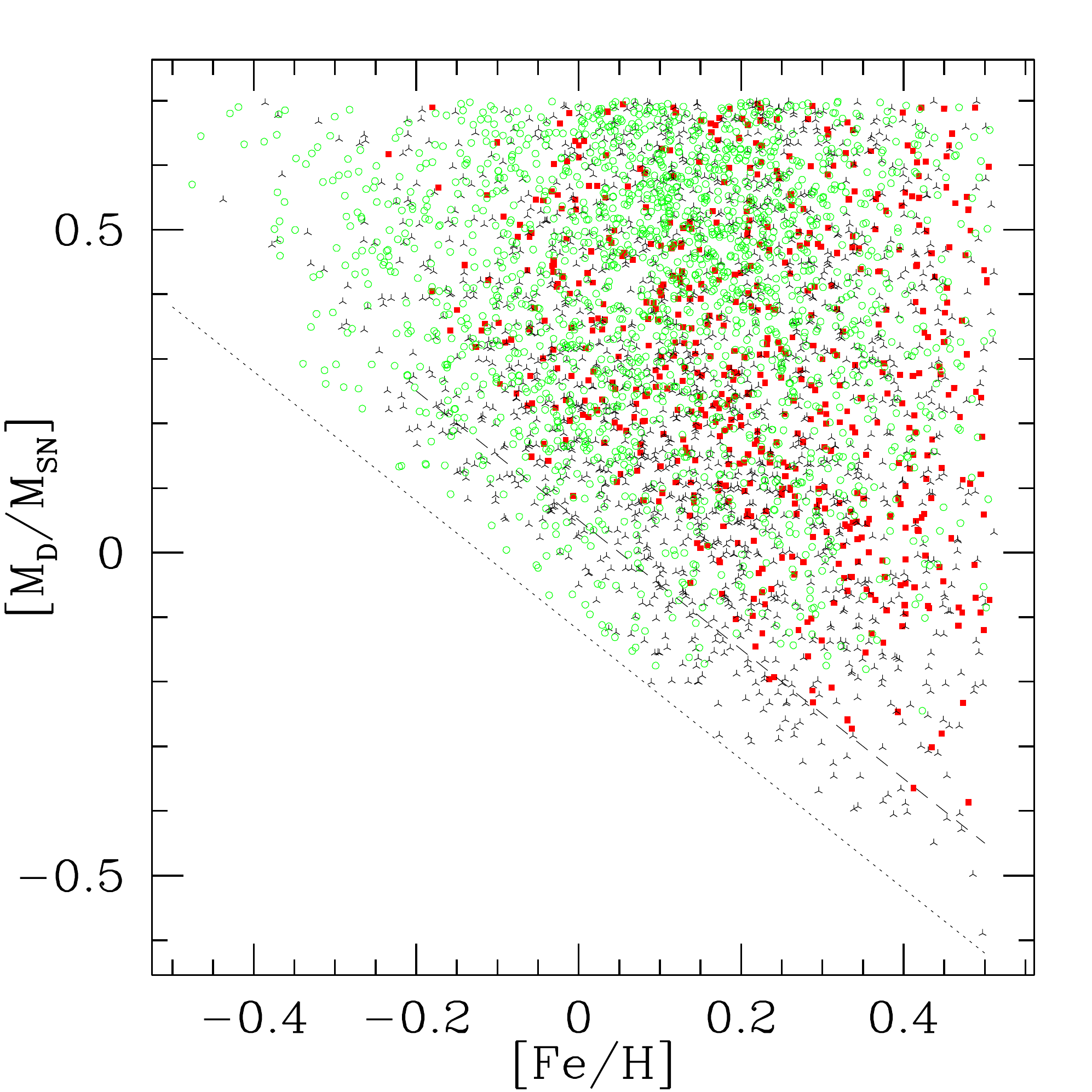}
\caption{Plane of initial conditions [Fe/H]-$\mm$ leading to the formation of synthetic planets larger than 300 $\mearth$. Different symbols indicate disk lifetimes: Red filled squares are low $\tdisk<2$ Myr. Black triangles are intermediate $\tdisk$ (2 to 4 Myr), and green open circles are disks with a long lifetime of 4 to 7 Myrs. The dotted line corresponds to an initial solid surface density of 6 g/cm$^{2}$ at 5.2 AU. Between the dashed and the dotted line, no short living disk can lead to giant planet formation.}\label{fig:fehmdplain} 
\end{figure}

Figure \ref{fig:fehmdplain} illustrates several compensation effects in the metallicity versus gas mass plane, with disk lifetimes coded with different symbols. Each point in the graph corresponds to an initial condition which allows the formation of a giant planet ($M >300\mearth$). Therefore, this plot shows three dimensions of the four dimensional parameter space of initial conditions  ($\astart$ is here projected into the plane).  If we would plot all initial conditions of the population, the full area of the figure would be filled with points, distributed as a two dimensional gaussian distribution with means just below zero dex both for [Fe/H] (-0.02) and $\mm$ (-0.04). 

The figure shows that only initial conditions lying in the upper right corner allow the formation of giant planets. The diagonal line corresponds to a solid surface density of $\sigmas=\fpg\sigmanorm=$6 g/cm$^{2}$, the minimal necessary value identified in Fig. \ref{fig:diskcondgiantsastartsd}. The fact that the points nearly fill the area above the line show that in this region, any combination of  the dust-to-gas ratio and the disk gas mass allows giant planet formation, provided that the product of these two quantities gives a value above the threshold value. This means, a low $\mm$ can be compensated by a high [Fe/H], and vice versa. A certain deviation of this purely multiplicative behavior can be seen at very low [Fe/H]. There, the mechanism plays that the very high necessary $\mm$ come along with a so large $\aice$ and thus typically large starting position $\astart$ and in turn so long core formation timescale, that the necessary $\sigmas$ increases stronger than in the rest of the parameter space, as discussed earlier in Sect. \ref{maximalmassesandmetallicity}. 

\subsubsection{Comparison with observations}
It is interesting to compare this result with the works of Greaves et al. (\cite{greavesetal2007}) and Wyatt et al. (\cite{wyattetal2007}). These authors derived, purely based on observational data (metallicities of giant planet hosts and dust disk mass distribution measured by submillimeter observations) that a minimal mass of solid elements of $0.5\pm0.1\,\mj$ in a disk is necessary form giant planets. In the model, which is in contrast built on first theoretical principles a minimal necessary $\sigmas$ of 6 g/cm$^{2}$ corresponds to an initial mass of planetesimals in the computational disk of about 0.4 to 0.5 $\mj$ (the spread comes from the position of the ice line which is not fixed by the product $\fpg\sigmanorm$). This is in excellent agreement with the observed value. In the model, there is no  threshold solid disk mass of any kind built in. In our models, and in contrast to Ida \& Lin (\cite{idalin2004a,idalin2004b,idalin2005,idalin2008}), not even a minimal core mass for the start of runaway gas accretion is built into the model. This threshold mass is not specified, but obtained by the solution of the planetary structure equations.  This agreement is therefore a good indication that our core accretion model catches important mechanisms occurring in giant planet formation. 

\subsubsection{Long necessary $\tdisk$ at low solid surface densities}\label{sect:longnecessarytdisk}
The colors and shapes of the symbols in Fig. \ref{fig:fehmdplain} show that there is another correlation between the initial conditions, explicitly not discussed by Greaves et al. (\cite{greavesetal2007}): The absence of red filled squares (showing short $\tdisk$) roughly between the dashed and  the dotted line shows that near the solid surface density threshold, only long living disks can produce giant planets. This is immediately understandable within the core accretion model: Near the threshold value of $\sigmas$, core formation is slower than at higher solid surface densities, which can be compensated by  long disk lifetimes to still be able to form giant planets.  Note that the distributions of disk lifetimes of our synthetic disks is by construction in agreement with the observed distribution (Paper I). It is interesting to note further that if the solar nebula indeed had a [Fe/H] and $\mm$ close to zero dex, we can deduce from Fig.  \ref{fig:fehmdplain}, that the solar nebula had a rather long disk lifetime of probably more than roughly 3-4 Myr, necessary to from Jupiter in such conditions. This in turn is  in good agreement with constraints coming from completely different measurement, namely that ages of CB chondrules and CV CAI which indicate a minimum lifetime of the solar nebula of $4.5\pm0.8$ Myr (Scott \cite{scott2006}). 

\section{Semimajor axis}\label{sect:semimajoraxis}
In the last section, we mainly studied the effects of disk properties on planetary mass. Another important observable quantity is the final  semimajor axis.  As explained in Paper II, one should keep in mind that the results concerning the semimajor axis are likely less firm than those for the mass. The reason is first that the type I migration model we use here is still based on the isothermal Tanaka et al. (\cite{tanakaetal2002}) migration rates, which were the only existing analytical description of type I at the time the model was written. Second we do not include any dynamical effects between multiple embryos. Nevertheless, we can gain insights of the role of several disk properties on planetary orbits in the idealized case of a single (massive) planet growing per disk in the limit of nearly absent type I migration.

\subsection{Extent of migration as function of  disk mass and [Fe/H]}\label{sect:extentofmigration}
Figure \ref{fig:ssodaofeho} shows as a function of disk gas mass (or equivalently, of the gas surface density) how far giant planets-to-be (final planetary mass larger than 300 $\mearth$) migrate, i.e. the difference $\Delta a$ between the final semimajor axis of the planet $a$ and the initial semimajor axis $\astart$ where its seed embryo is put into the disk. Both type I and type II migration rates for the planet dominated regime, which is the most relevant one (Paper I), are proportional to the gas surface density. Note that by construction, only inward migration is possible in the current model (Alibert et al. \cite{alibertetal2005a}): For type I migration, the equations of Tanaka et al. (\cite{tanakaetal2002}) in practice always yield negative torques, and for type II we implicitly assume to be well inside the radius of maximum viscous couple, which is justified in almost all circumstances (Paper I). The colors and symbols additionally show the corresponding metallicity of the disk. A quite complex pattern is seen, which is the consequence of several effects:

\begin{figure}
\includegraphics[width=\columnwidth,origin=left]{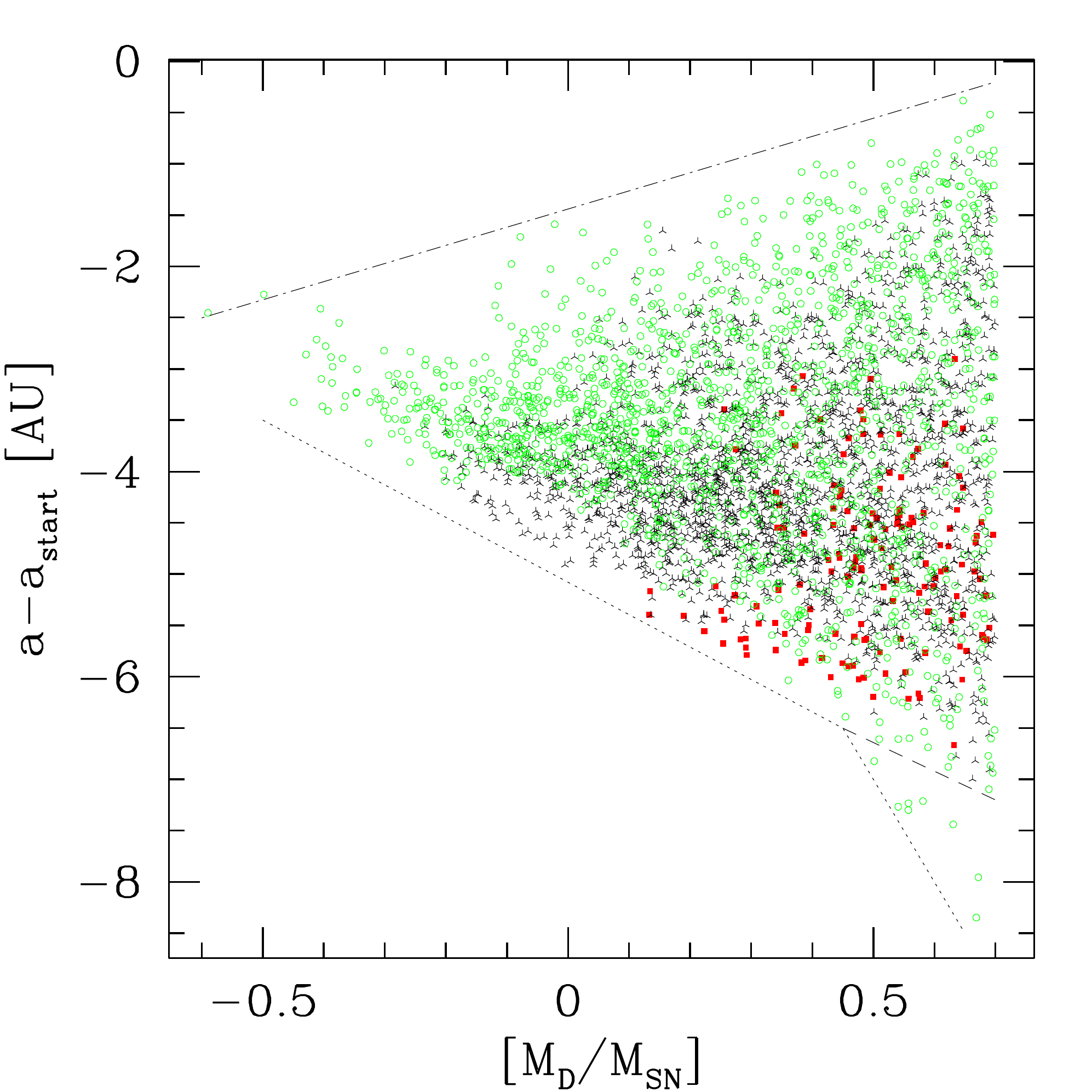}
\caption{Extent of migration $\Delta a=a-\astart$ as a function of the disk gas mass $\mm$ for synthetic planets eventually becoming giant planets ($M\geq300\, \mearth)$. Symbols indicate disk metallicities:  Red filled squares are low [Fe/H]$<-0.2$. Black triangles are intermediate [Fe/H] (-0.2 to 0.2 dex), and green open circles are disks with a high [Fe/H]$>0.2$. A number of lines are plotted to show relevant limiting regimes (see text).}\label{fig:ssodaofeho} 
\end{figure}

One expects that at a high disk gas mass, the largest extent of migration happens. This can indeed be the case, as indicated by the dotted line.   At the highest  $\mm$, the extent of migration can be very large, up to -8 AU as seen in the lower right corner of the graph. But it is more complex: when the disk mass is high, the extent of migration can also be minimal as shown by the green points in the upper right corner, where only about 0.5 AU of radial displacement occurs. The difference comes from the metallicity, as shown by the symbols. In the bottom right corner, there are a many red squares, indicating low  [Fe/H]$<-0.2$, which do not exist in the top right corner. (Ignore for the moment the few green open circles found in the outermost bottom right corner between the dashed and the dotted line. The very large displacement $\sim -8$ AU there are special cases addressed further down). 

\subsubsection{Braking effect}\label{sect:brakingeffect}
The reason is that metallicity first acts through a ``braking effect'': When both the gas mass and the metallicity is high, the planetesimals surface density is high. Therefore, even if the torques would be strong, planets get quickly so massive (undergo quickly runaway gas accretion) that the disk soon cannot force them to migrate on the viscous timescale any more, as the planet mass is larger than the local disk mass (Lin \& Papaloizou \cite{linpapaloizou1986}). Rather, these planets quickly get into the slower planet dominated type II migration (Armitage \cite{armitage2007}) where the migration rate is inversely proportional to the planet mass. Therefore, if both the gas mass and the metallicity is high, planets do not migrate much if the planetary embryo starts at a rather small semimajor axis (which is possible in metal rich disk as we have seen above) because there the local disk mass is small (Paper I).  We see thus that via the planet mass, there exists a negative correlation between metallicity and migration, even if no influence of the metallicity on the migration rate is directly included in the model. 

\subsubsection{Collection effect}\label{sect:collectioneffect}
Second, metallicity acts through a ``collection effect'': When the gas mass is high, but [Fe/H] is low, then the extent of migration is large (red squares in the bottom right corner). Under such disk conditions, the planetesimal surface density is only intermediate, but the torques are strong. Cores cannot become supercritical for gas runaway accretion in situ. In order to become a giant planet, they must first migrate inwards through the disk, collecting the planetesimals they come across, until they reach a mass where gas runaway accretion sets in (see the formation tracks in Paper I).  Then the planet slows down due to the mentioned ``braking effect''. In other words, a lower planetesimal surface density and a high gas mass make that planets migrate strongly, because they stay longer in the faster disk dominated type II migration. Therefore, the extend of migration is large, of order -4 to -6 AU. The collection effect makes that the minimal extent of migration (upper limit of the envelope indicated by the dashed-dotted line) increases with decreasing disk mass. Note that the ``braking'' and the ``collection'' effect mean that accretion and migration are two interdependent mechanisms that should not be treated independently. 

Third, one can see in the graph again the ``compensation effect'' discussed above in Sect. \ref{sect:compensationeffects}, i.e. the fact that at low gas masses, high [Fe/H] are necessary to form a giant planet. This is illustrated by the absence of red squares at low $\mm$. Instead, if the gas surface density is low, [Fe/H] must be high (only green and black symbols). At these rather low disk masses, migration is of intermediate importance, as the torques are not very strong, but at the same time the cores have to migrate over a certain distance until they have collected enough solids to get into gas runaway accretion, and slow down. Therefore, the extend of migration is intermediate at low $\mm$, about -3 AU.

Fourth, we finally also notice a consequence of the ``timescale effect''  mentioned before in Sect. \ref{sect:isomassandtimescale}.  The fact that in the outermost lower right corner (between the dashed and the dotted line) the symbols are  all green circles only is the consequence of the following: to be able to migrate over such a large distance, planets must already start at large initial semimajor axes and nevertheless manage to grow supercritical on a timescale shorter than the disk lifetime. This is however only possible for special cases when the solid surface density is very high, otherwise the formation timescale for a massive core is too large. Therefore, the planets that migrate most are again found in disks with a large mass and [Fe/H], and with a large $\astart$. They don't stop due to the ``braking effect'', as the local disk mass which is given approximately as $\Sigma a^{2}$ is large at large distances.

\subsubsection{Dependence on model assumptions}
We have studied how these results depend on model assumptions, by looking at several non-nominal populations. For a population where the ice line is fixed at 2.7 AU independently of $\mm$, the result remains quite similar.

In a non-nominal population where type II migration is only partially suppressed in the planet dominated regime, i.e. where the reduction of the type II migration rate due to the planet's inertia is assumed to scale  proportional to  $(\Sigma a^{2}/\mplanet)^{1/2}$ instead of linearly (cf. Alexander \& Armitage \cite{alexanderarmitage2009}), the triangular envelope shape of the points is approximately retained, but shifted to larger extents of migration. For example, the extent of migration at the lowest $\mm$ increases from -3 to about -5 AU, and the maximal overall distance is about -15 instead of -8 AU. 

Using a type I migration efficiency factor of $\f1=0.1$ has a remarkably weak influence on the dependence of $\Delta a$ on the disk properties,  except from shifting the whole envelope of points to larger values by about -0.5 AU. 

The plot also indicates that for solar nebula like conditions ([Fe/H] and $\mm\approx0$), a $\Delta a$ of about -4 AU is found, compatible with our earlier results for the formation of Jupiter (Alibert et al. \cite{alibertetal2005b}).

\begin{figure}
\includegraphics[width=\columnwidth,origin=left]{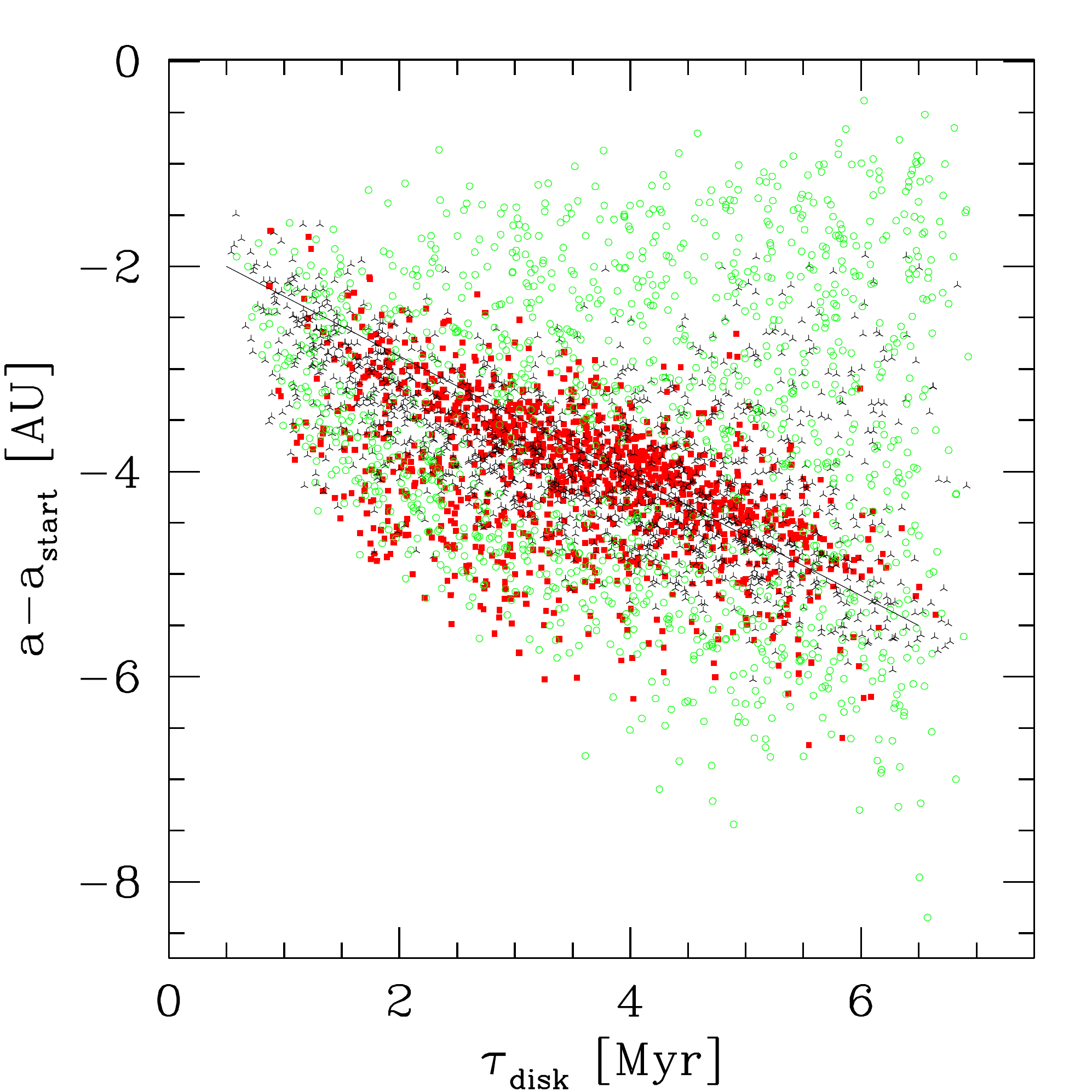}
\caption{Extent of migration $\Delta a=a-\astart$ as a function of the disk life time $\tau_{\rm disk}$ for synthetic planets eventually ending up as giant planets ($M\geq300\, \mearth)$. Symbols indicate relative surface densities of planetesimals  $\sigmas/\sigmassn$:  Red filled squares are  low $\sigmas/\sigmassn<2$. Black triangles are intermediate  $\sigmas/\sigmassn$ (2 to 4), and green open circles are disks with a high  $\sigmas/\sigmassn>4$. The solid line indicates the very roughly linear behavior of $\Delta a$ as a function of   $\tau_{\rm disk}$. There are however large systematic departures  from this simple dependence. }\label{fig:tdiskodaossso} 
\end{figure}

\subsection{Extent of migration as function of $\tdisk$}
In figure \ref{fig:tdiskodaossso}  we have plotted the extent of migration as a function of the life time of the protoplanetary disk. The symbols indicate the relative amount of solids in the disk, $\sigmas/\sigmassn$, cf. Fig. \ref{fig:diskcondgiantsastartsd}. The boundaries for the three bins were chosen in a way that each of the three bins contains about the same number of planets. As expected (Alexander \& Armitage \cite{alexanderarmitage2009}) disks with a long lifetime drive migration over larger radial extents. The plot shows that the bulk of the planets roughly lies  on a line (indicated in the plot) running from $\Delta a\approx -2$ AU for the disk with the shortest life time which still can form a Jupiter mass planet (about 0.5-1 Myr) to a  $\Delta a\approx -5.5$ AU for the disks with the longest $\tau_{\rm disk}$ of about 7 Myrs. Such a relationship is expected, because there is first a characteristic locus from where giant planets come (sect. \ref{sect:startingposition}), and second the longer the disk lifetime, the longer the torques can act to move the planet. Cases falling on the line mostly come from the intermediate bin of   $2\leq\sigmas/\sigmassn\leq4$, which corresponds to typical values of the solid surface density. 

\begin{figure*}
\begin{minipage}[lt]{0.5\textwidth}
\includegraphics[width=\textwidth,angle=0]{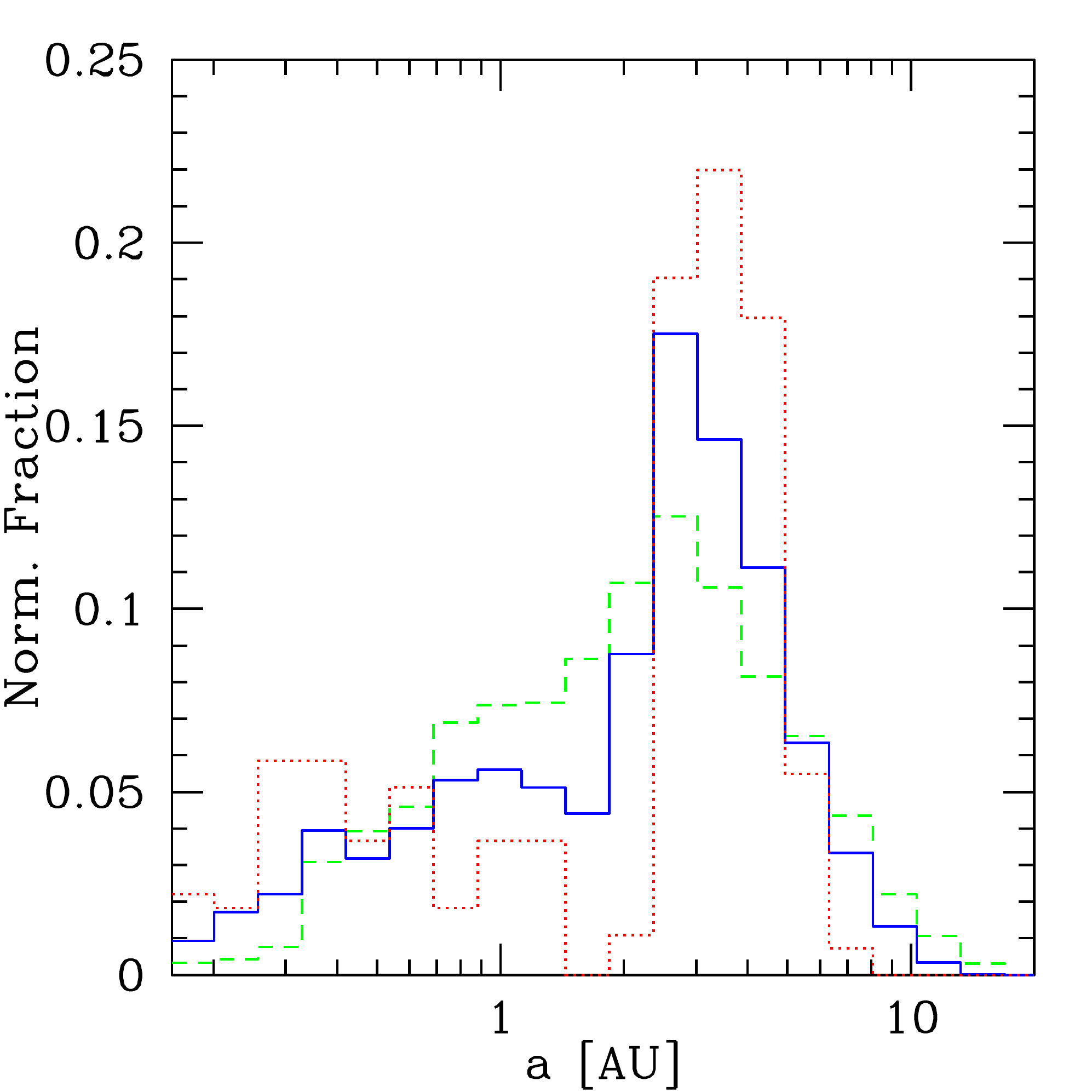}
\end{minipage}
\hfill
\begin{minipage}[lt]{0.5\textwidth}
\includegraphics[width=\textwidth,angle=0]{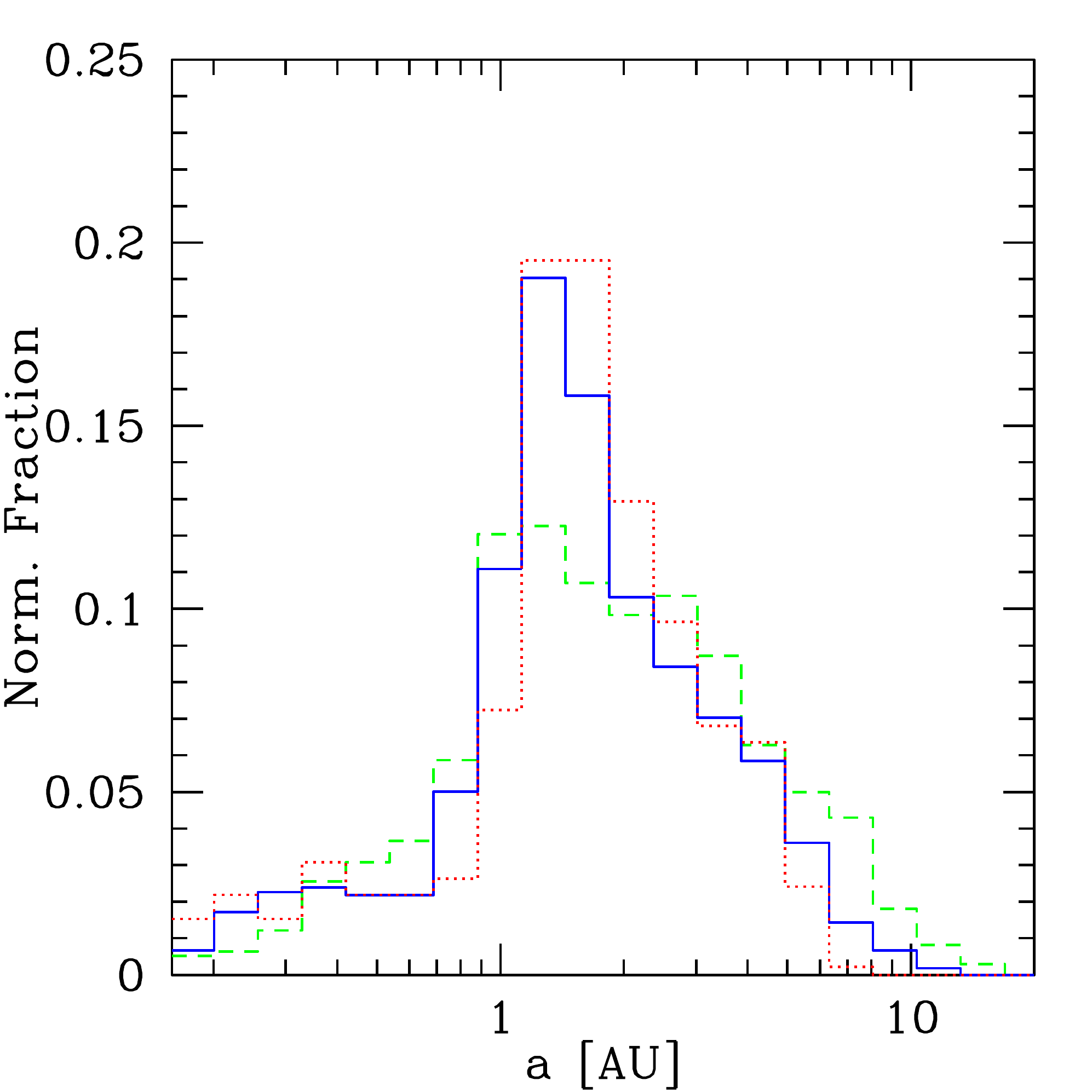}
\end{minipage}
\caption{Distribution of final semimajor axes of synthetic planets larger than $300\,\mearth$, as a function of metallicity. In both panels, the red dotted lines are for [Fe/H]$<-0.2$, the blue solid lines for $-0.2<$[Fe/H]$<0.2$, and green dashed lines are for [Fe/H]$>0.2$. The left panel is the nominal population, while the right panel is a population where $\aice=2.7$ AU fixed for all initial conditions. Hot planets ($a\lesssim0.1$ AU) are not shown in this figure.}\label{fig:afhistofeh}
\end{figure*}

There is however a significant spread around the diagonal line. There are, for example at each $\tau_{\rm disk}$ points which lie above the plotted line, i.e. where only very little migration occurs. The symbols  in this area (green circles and black triangles) indicate that these are disks with a high solid surface density. One then finds that such cases all correspond to planets starting inside the ice line, which is possible for high $\sigmas/\sigmassn\gtrsim3.5$, see Fig.  \ref{fig:diskcondgiantsastartsd}. As explained in the previous section, these planets don't migrate much, because of the breaking effect that sets in soon. 

On the other hand, there are  also cases where clearly more migration happens than for the bulk of the planets, i.e. which lie clearly below the line. The symbols indicate that this is a mixture of both low (red squares) and high (green circles) solid surface densities. The former class corresponds to planets that start not far outside the ice line, but then migrate all the way down close to the star (0.1-0.7 AU), because of the ``collection effect'', whereas the later correspond to planets starting at very large semimajor axes, well outside the ice line, which is possible in such metal rich disks (cf.  Fig. \ref{fig:daiceastart}), and migrate then a lot because  of the braking effect setting in late at such distances, but still never end up inside $\sim 3$ AU.

\subsection{Final semimajor axis distribution of giant planets}\label{sect:semimajoraxisdist}
As mentioned in the introduction, there is no clear imprint of [Fe/H] onto the semimajor axis distribution of giant planets in the observational data, see e.g. Valenti \& Fischer, (\cite{valentifischer2008}) or Udry \& Santos (\cite{udrysantos2007}).  It was studied repeatedly (e.g. Sozetti \cite{sozzetti2004}; Santos et al. \cite{santosetal2006}) whether stars hosting a Hot Jupiter are particularly metal rich, even among exoplanet hosts, manifesting in an absence of Hot Jupiter host stars with a clear subsolar metallicity. A number of  Hot Jupiters around at least fairly low [Fe/H] $\sim-0.2$ to -0.4 hosts are now known, and statistical tests point to rather insignificant differences in the metallicity distribution of close-in planets and those on wider orbits (Ammler-von Eiff et al. \cite{ammlervoneiffetal2009}). On the other hand, planets of metal poor stars might have a  tendency to have rather large semimajor axes outside 1.5 AU (Santos et al. \cite{santosetal2009}). The basic problem is that our samples are still too small for definitive conclusions.

The lack of a clear imprint of [Fe/H] on the semimajor axis might seem surprising at first glance from a theoretical point of view seen the significant influence of [Fe/H] on $\astart$ or $\triangle a$ discussed in the previous chapters. However, as we will demonstrate below, once these different effects are combined, one can actually understand such a weak dependance  well on theoretical grounds.

Figure \ref{fig:afhistofeh} shows the distribution of final semimajor axes  for planets more massive than 300 $\mearth$. The planets are binned into low, medium and high [Fe/H] cases. Note that each metallicity bin was normalized individually, so the absolute heights cannot be compared between different metallicity bins. The plot shows the nominal population (left panel), and a population with $\aice=2.7$ AU fixed (right panel). Note also that the ``hot'' planets, i.e. those planets that reach the inner boarder of the computational disk at about 0.1 AU are not included here. Due to the curvature of the feeding limit, also the bins out to a distance of about 0.2 AU are artificially lowered (see Fig. \ref{fig:aM4p}). The planets in the feeding limit are addressed below (Sect. \ref{sect:metallicitycloseinplanets}). 

While the exact numerical values are different for the two populations, we see that the general shape of the semimajor axis distribution is similar, and consists first of a slow increase of the planet frequency with increasing semimajor axis for high [Fe/H] respectively an approximately flat part for lower [Fe/H], then an upturn to a maximum at a few AU (at 2-3 AU on the left, at $\sim1$ AU on the right), followed by a gradual decline at even larger radii. This is due to the fact that the core accretion timescale becomes too long at large distances (Sect. \ref{sect:isomassandtimescale}). Our model does not include at the moment mechanisms like scattering that could bring planets to larger semimajor axes. This probably leads to an underestimate of the number of giant planets at large distances, which is important for direct imaging searches (e.g. Bonavita et al. \cite{bonavitaetal2009}). This issue will be addressed in forthcoming work. 

\subsubsection{Imprint of the iceline}
For the nominal population, the location of the peak can roughly be understood by considering some relevant mean values (median values give similar results) for the planets with a final mass $>300\,\mearth$: First, the mean $\sigmanorm$ of these planets is about 480 g/cm$^{2}$, corresponding to an $\aice\approx5.5$ AU. As indicated by Fig. \ref{fig:daiceastart}, the mean starting position should be somewhat outside of this distance. The measured mean $\astart$ is indeed 6.5 AU. The mean disk lifetime of 3.8 Myr corresponds to a mean  $\Delta a\approx -3.8$ AU (Fig. \ref{fig:tdiskodaossso}), so that we estimate a mean final semimajor axis of 2.7 AU, which is the same as the mean value found in the simulations. This corresponds approximately to the peak of the blue line, which is the bin with the highest number of planets in it. Thus, the peak of the distribution is due to the fact that there is a typical locus from where planets come, combined with a typical distance over which they migrate. 

This imprint of the ice line into the semimajor axis distribution is confirmed by the right panel, where $\aice$ is at 2.7 AU for all disks. Compared to the nominal case, the upturn in the distribution is found at a smaller semimajor axis, namely at about 1 AU which might be closer to the observed distribution and might also be sharper, at least for the high metallicity bin. These differences can be understood: The sharpness is a logical consequence because part of the spread in the typical $\astart\gtrsim\aice$ is now simply eliminated by suppressing the dependence of $\aice$ on $\sigmanorm$. The fact that the upturn happens at a smaller distance simply comes from the fact that  an $\aice=2.7$ AU is clearly less that the mean $\aice$ in the nominal case (5.5 AU).  We conclude that there is a correlation between the position and shape of the upturn in the planetary semimajor axis distribution, and the thermodynamic properties in the protoplanetary disk, which is an interesting result. 

\subsubsection{Comparison with observations and predicted decrease}
Note that we here only consider the condensation of ices as a mechanism causing a sweet spot for giant planet formation. Other (or additional) mechanism with comparable consequences could also be at work in the disk (cf.  Schlaufman et al. \cite{schlaufmanetal2009}). One should also keep in mind that the temperature and solid surface density structure of protoplanetary disks  is likely much more complex than assumed in the simple $\alpha$ models used here (Dzyurkevich et al. \cite{dzyurkevichflock2010}).  This means that the result that the frequency of giant planets is an increasing function of the semimajor axis inside a few AU is a more robust prediction than the specific shape and cause of it. Observationally, after correction of the observational bias, an overall distribution of the semimajor axes of giant planets not dissimilar from the one here has been found recently (Mayor et al. \cite{mayormarmier2011}). No imprint of [Fe/H] is seen in the still small dataset, where no fine structures (like e.g. the small valley visible at low [Fe/H] just inside the upturn) could currently be seen. It should be noted that such fine structures are sensitive to specific model assumptions.

The baseline of the RV surveys is also not yet long enough to determine if the theoretically predicted decrease at  larger distances  ($\gtrsim5$ AU) exists. Such a decrease can in contrast be seen as a solid prediction of the core accretion theory. The observational test of this will be an important constraint for the core accretion model, as it is linked to the core growth timescale. A complication arises from the fact that the gravitational interaction of several giant planets can lead to the scattering of bodies to larger semimajor axes. This evolutionary effect will blur the timescale limit set during formation. To still be able to use the observational constraint, formation and evolution through N-body interaction must be coupled self-consistently (Alibert et al. in prep.).

\subsubsection{Independence from [Fe/H]}\label{sect:independencefromfeh}
Comparing the three metallicity bins in the nominal case, we see that there are some visible, but not extreme differences. For the low metallicity bin, one sees that this distribution is rather flat inside about 3 AU, then has a sudden upturn, followed again by a sharp downturn. Thus, the planets are mostly well confined to a particular region. For the high metallicity bin, and to a lesser extent also for the intermediate metallicities, the increase at small semimajor axes is more gradual towards the peak, and also the decrease towards the even larger semimajor axis is less abrupt, which means that the high [Fe/H] distribution has more planets both at smaller and larger distances than the low metallicity one.

These finding are due to the fact that giant planets around metal poor stars can only form within a well defined region, namely at the sweet spot for giant planet formation as was shown in Sect. \ref{sect:startingposition}, in contrast to higher [Fe/H], which can form all over the disk. Considering the location of the peak for the low metallicity bin, one notes that the peak of giant planets around low metallicity stars is found somewhat further out by (about 1 to 3 AU) than for the higher metallicity cases. This might be in accordance with the observational findings of  Santos et al. (\cite{santosetal2009}).  This can also be understood in terms of our findings from the previous sections: At low metallicities, giant planets must come from outside the ice line (Sect. \ref{sect:startingposition}), i.e. from a significant distance. At the largest semimajor axes ($a\gtrsim6$ AU), the relative frequency of more metal rich planets is again higher.  This is a consequence of the timescale effect: a lot of solids are necessary to form a core in time at large distances, and this is not the case at  low [Fe/H].

These reasonings alone are however not yet sufficient to explain the observed (and theoretical) semimajor axis distribution with its weak [Fe/H] dependence, as it would rather imply a strong absence of giant planets around low metallicity stars at small semimajor axes. The missing link is the anti-correlation of metallicity and the extent of migration (sect. \ref{sect:extentofmigration}), due to the ``braking effect'' which stops migration quicker in metal rich disk, and the ``collection effect'' which enhances it in solid poor ones. Thus, while giant planets initially arise closer-in in high [Fe/H] environments, they migrate less, while for low [Fe/H], they start further out, but migrate more. 

These findings can be quantified  (Table \ref{tab:meana}): For the nominal population, the mean $\astart$ is 6.2, 6.7 and 7.5 AU for the high, medium and low [Fe/H] group. But the mean $\Delta a$ is -3.5, -4.0 and -4.9 AU again for high, medium and low [Fe/H], which results in a near cancelation of the two counteracting effects, so that the resulting mean final semimajor axis for the three metallicity bins are all the same except for a tiny difference of 0.1 AU. Observationally,  such three mean values appear indistinguishable.  Thus, a complex of interplay of migration and accretion explains the observed absence of a strong correlation of metallicity and semimajor axis for extrasolar giant planets.

\begin{table}
\caption{Mean values of the starting position $\astart$, the extent of migration $\Delta a$, and the final semimajor axis $a$ for planets with a final mass larger than 300 $\mearth$,  as a function of metallicity. Planets inside 0.1 AU are excluded.}\label{tab:meana}
\begin{center}
\begin{tabular}{lccc}
\hline\hline
  Metallicity                        & $\astart$  [AU]         & $\Delta a$ [AU]    & $a$ [AU] \\ \hline      
[Fe/H]$>0.2$               &  6.2            & -3.5  &  2.7 \\ 
$-0.2<$[Fe/H]$<0.2$    & 6.7              &  -4.0            &  2.7  \\
$$ [Fe/H]$<-0.2$    & 7.5             &  -4.9            &  2.6  \\ \hline
\end{tabular}
\end{center}
\end{table}

\section{Metallicity of close in Jovian and Neptunian planets}\label{sect:metallicitycloseinplanets}
\subsection{Observed distribution}

\begin{figure*}
\begin{minipage}[lt]{0.666\textwidth}
\includegraphics[width=\textwidth]{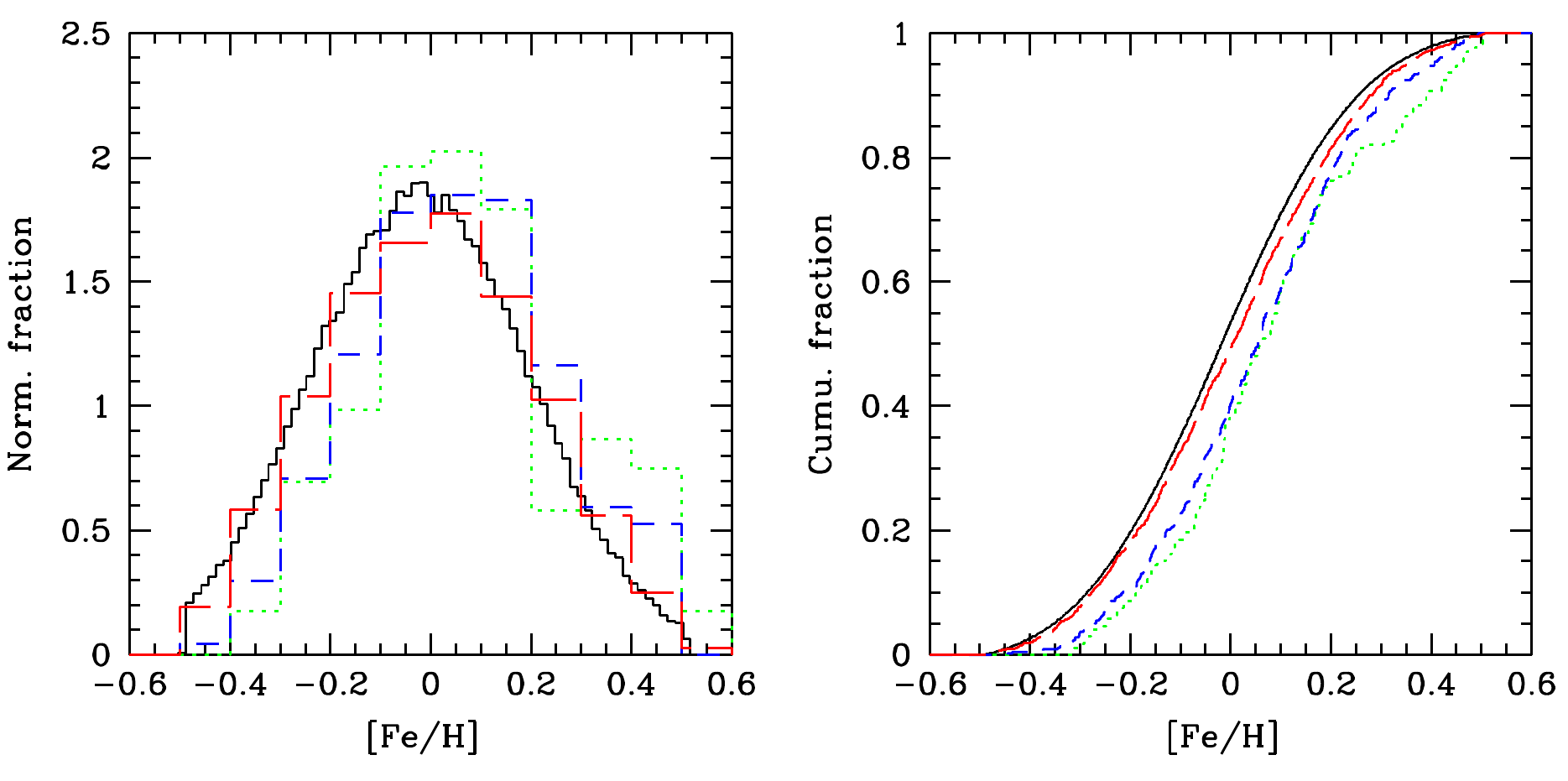}
\includegraphics[width=\textwidth]{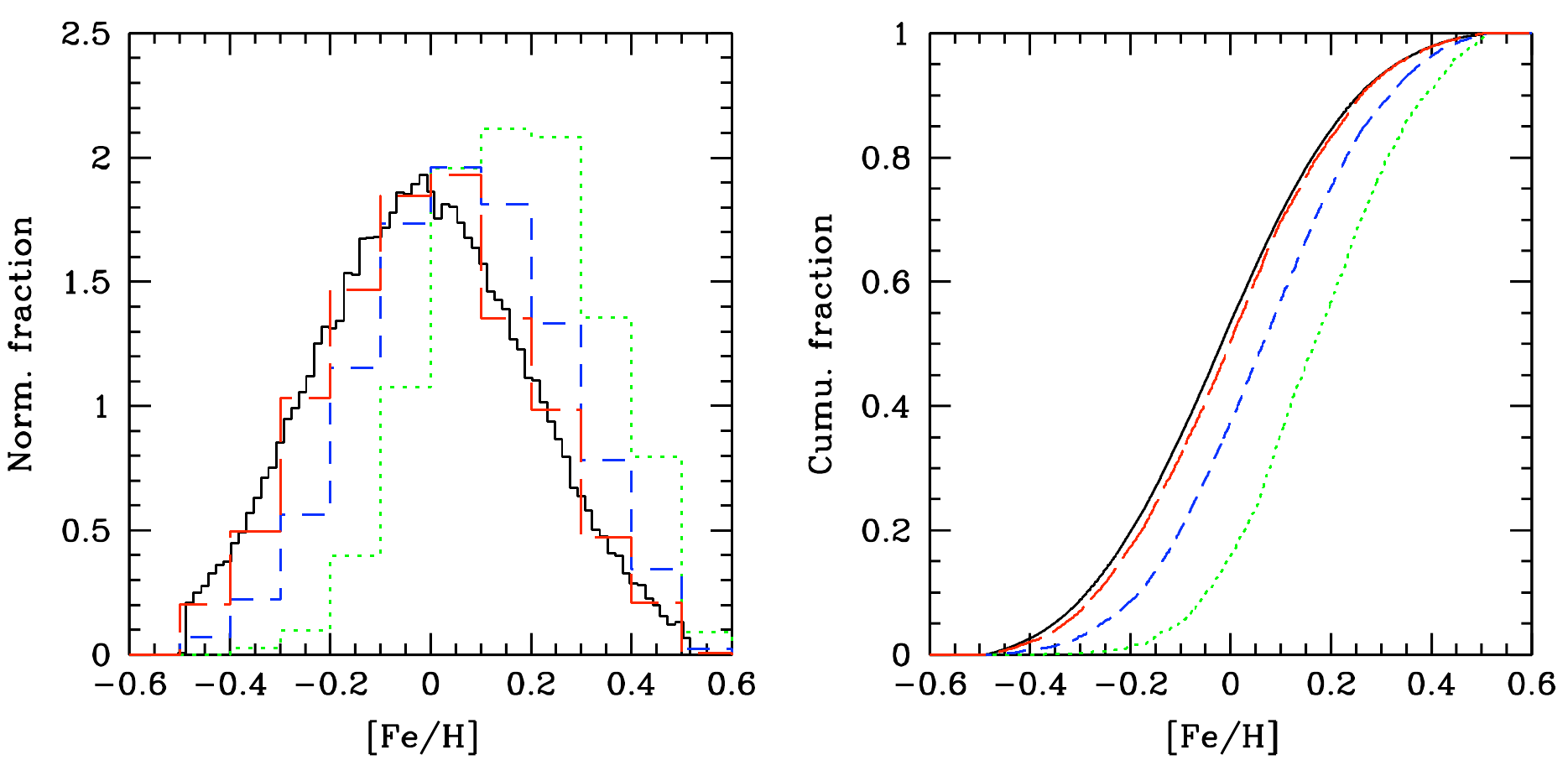}
\end{minipage}
\hfill
\begin{minipage}[lt]{0.33\textwidth}
\caption{Histogram and cumulative distribution of the metallicity of close-in synthetic planets. The upper two panels are for the nominal population ($\f1=0.001$), the lower two for a population with a faster type I migration ($\f1=0.1$). In all panels, green lines stand for massive planets ($M>100\,\mearth$), the blue lines are planets with $100>M/\mearth>6\,\mearth$,  and red lines are low mass planets ($M/\mearth<6$), all inside the feeding limit, i.e. inside about 0.1 AU. The black lines serves for comparison and shows the metallicity distribution of all synthetic planets (at all semimajor axes) and thus simply the distribution from which we draw the initial conditions.}\label{fig:fehhotdist}  
 \end{minipage}
 \end{figure*}

Early detections made with high precision radial velocity observations (Udry et al. \cite{udryetal2006}) suggested that the distribution of the host star metallicities of close-in lower mass Neptunian and Super-Earth planets follows a different pattern than observed for hot Jupiters, namely that their metallicity distribution is not shifted towards the higher metallicities. This result was subsequently discussed in several studies (e.g. Sousa et al. \cite{sousaetal2008}; Ghezzi et al. \cite{ghezzietal2010}) and recently confirmed (Mayor et al. \cite{mayormarmier2011}; Sousa et al. \cite{sousaetal2011}).

The latter authors find that the mean metallicity of  FGK hosts stars is positively correlated with the mass of their most massive planet. They find a mean  [Fe/H] of -0.11,  0.04 and 0.10 dex for host stars having planets with masses between 0.01 and 0.1 $\mj$, 0.1 and 1 $\mj$ and $>1 \mj$, respectively.  In a similar way show Ghezzi et al. (\cite{ghezzietal2010}) that for FGK host stars, in the observational data, there is an offset in the mean metallicity of +0.11 dex by which Jovian planet hosts are more metal rich than  stars which host Neptunian planets only. The probability that the two samples are drawn from the same parent distribution is however found to be 17\%, i.e. clearly non-negligible.  They give a mean metallicity for Neptunian planet only FGK hosts of 0.01 dex, and of 0.12 dex for Jovian planet FGK hosts.

\subsection{Synthetic result}
It is obvious that this raises the question whether such a relation exists for the synthetic population, too. From Sect. \ref{subsubsect:imf} such a behavior is expected, but the problem is that there, the mass function is shown for all semimajor axes, while the observed planets with  $M \leq 30\,\mearth$ are almost all inside of a few 0.1 AU.

We therefore plot in Fig. \ref{fig:fehhotdist} the metallicity distribution\footnote{With metallicity of a planet we refer in this work  always to the disk/stellar [Fe/H] of the initial conditions which lead to the formation of this planet, and not to the heavy element abundance in the planet itself.} of those synthetic planets that have migrated into the feeding limit at about 0.1 AU (``Hot'' planets), for three different mass bins: $M >100 \mearth$ (green lines), $100>M/\mearth>6$ (blue lines) and $M/\mearth<6$. The lower limit of 6 $\mearth$ approximately reflects the lower limit to which the model should yield masses which are not too much affected by the initial conditions (Paper I). Below this limit we can still conclude that there are additional low mass planets, but their final mass is not well described in the model. The upper two panels in Fig. \ref{fig:fehhotdist} show the nominal population with a type I efficiency factor $\f1=0.001$, while the two lower panels show a non-nominal population with $\f1=0.1$. Numerical values for the mean metallicity   of the different sub-populations are listed in table \ref{tab:fehhots}.
 
\begin{table}
\caption{Mean [Fe/H] of the initial conditions leading to different types of synthetic planets.}\label{tab:fehhots}
\begin{center}
\begin{tabular}{lcc}
\hline\hline
Mean [Fe/H]                       & Nominal           & $\f1=0.1$    \  \\ \hline                                                      
All planets (initial cond.)                                  & -0.02            & -0.02 \\ 
Hot, $M/\mearth<6$                  &  0.00            &  0.00  \\
Hot, $6<M/\mearth<100$                  &  0.05            &  0.06  \\
Hot, $M/\mearth>100$                  &  0.08           &  0.17  \\
\hline
\end{tabular}
\end{center}
\end{table}

In all figures, the black line additionally shows for comparison the [Fe/H] distribution of all synthetic planets, and thus the distribution from which the initial conditions are drawn. It is a Gaussian with a mean $\mu=-0.02$ dex and a dispersion $\sigma=0.22$ dex (cf. Paper I). 

 \begin{figure*}
\begin{minipage}[lt]{0.666\textwidth}
\begin{minipage}[lt]{0.5\textwidth}
\includegraphics[width=\textwidth,angle=0]{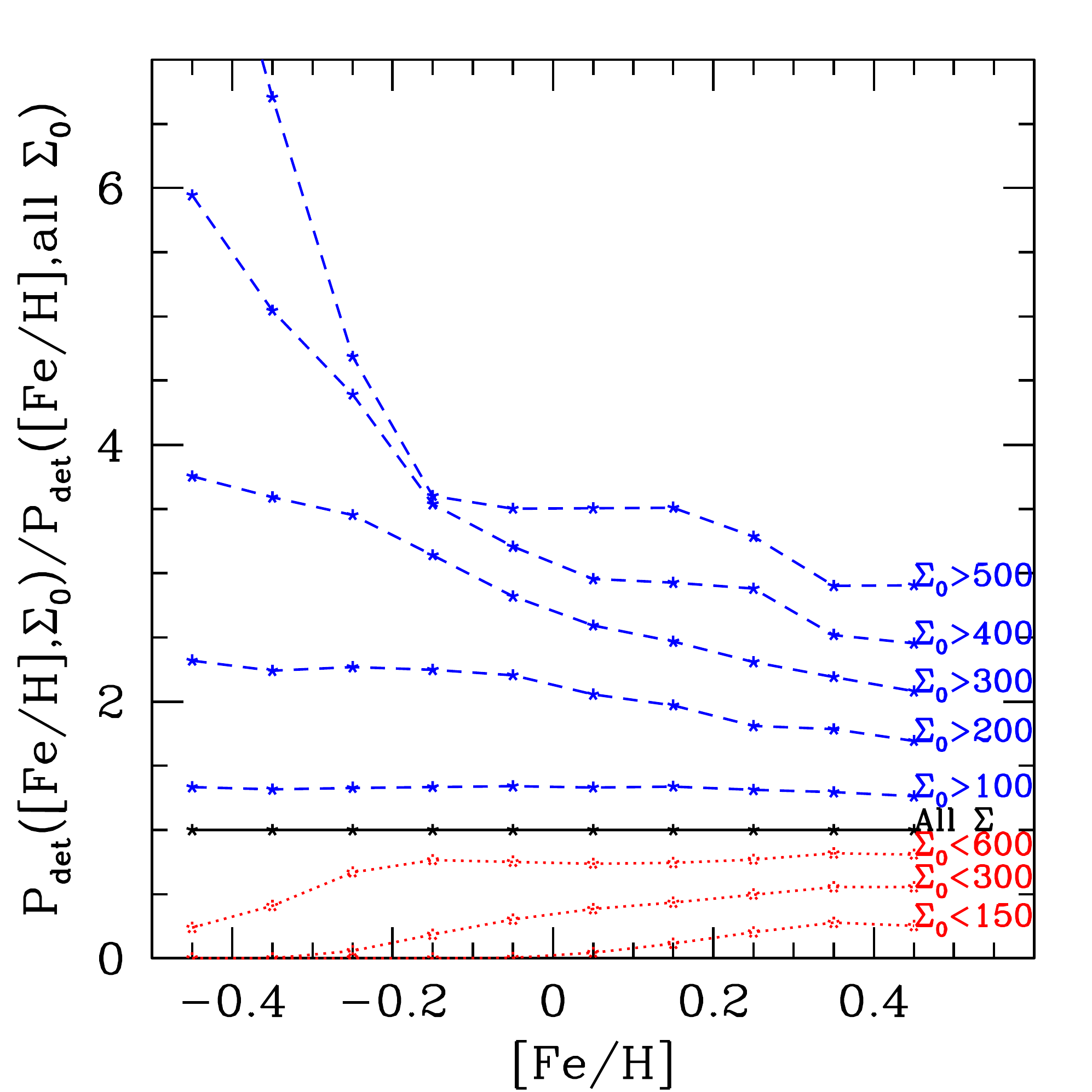}
\end{minipage}
\hfill
\begin{minipage}[lt]{0.5\textwidth}
\includegraphics[width=\textwidth,angle=0]{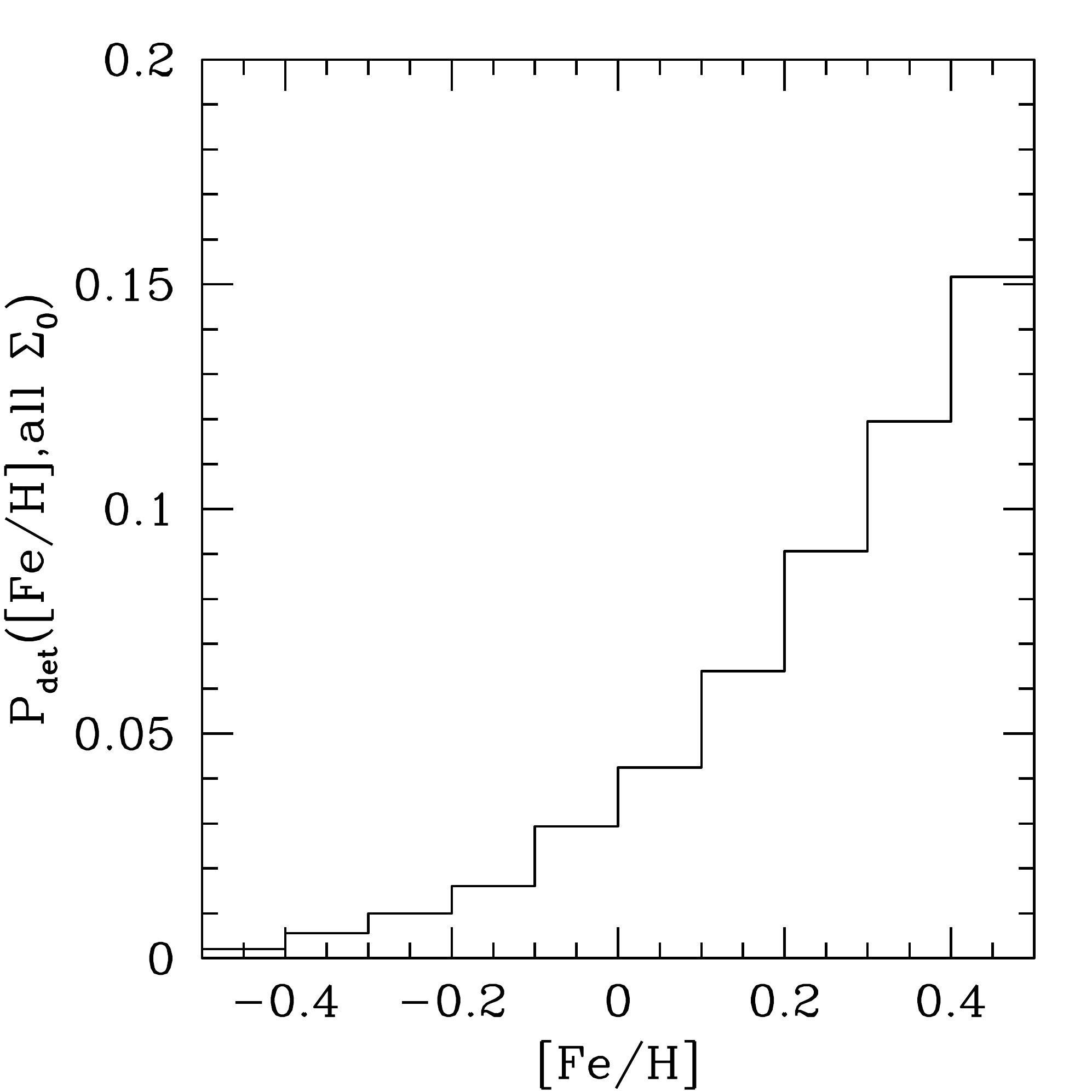}
\end{minipage}
\end{minipage}
\hfill
\begin{minipage}[lt]{0.33\textwidth}
\caption{Left panel: Detection probability as a function of metallicity  and gas surface density. The plot shows the relative detection probability as a function of [Fe/H] for disks which have a $\sigmanorm$ meeting the condition indicated in the plot, normalized by the $P_{\rm det}$([Fe/H], all $\sigmanorm$) i.e. the detection probability as a function of [Fe/H] alone, including all $\sigmanorm$. This quantity is plotted in the right panel. A curve in the left panel which increase towards the low [Fe/H] (blue lines) therefore means a less strong metallicity effect, while a curve that decreases towards the low [Fe/H] (red lines) indicates a stronger metallicity effect. The instrumental accuracy is  10 m/s.}\label{fig:pdetecfeh}
 \end{minipage}
 \end{figure*} 

One notes that for both populations, the hot Jovian planets have a distribution that is more metal rich than the intermediate mass and Neptunian planets or the Super-Earth planets. For the nominal distribution, the difference between the intermediate and the high mass bin is however extremely small, except for a possible excess of Hot Jupiters with a high metallicity $\gtrsim0.2$ dex. In terms of mean metallicities, the difference is just 0.03 dex. For the population with $\f1=0.1$, the situation is quantitatively quite different: Here, the mean metallicity of the Jovian mass bin is with 0.17 dex clearly higher (+0.11 dex) than the middle bin with the Neptunian planets (0.06 dex).  For the lowest mass bin, which contains the proto-terrestrial planets, a vanishingly small offset to the distribution from which the initial conditions are drawn is seen, for both synthetic populations, as these planets have a mean [Fe/H]=0.00.

We thus see, as expected, that the mean [Fe/H] increases with increasing planetary mass. Quantitative results are however dependent on poorly constrained model parameters like $\f1$. The large difference for the mean metallicity of Jovian planets for the two populations stems in particular from the following: Giant planets forming in high [Fe/H] and low $\sigmanorm$ environments inside the ice line do not usually migrate into the feeding limit provided that  type I migration is negligible ($\f1=0.001$), because of the very efficient braking effect in type II migration at such small semimajor axes (sect. \ref{sect:extentofmigration}). They rather stay at distances between roughly 0.4 and 1 AU. For $\f1=0.1$ this is different: there, type I migration, which has a migration rate that increases in contrast linearly with planet mass, brings the quickly growing cores starting inside the ice line already so close to 0.1 AU, that many planets forming in such a high [Fe/H] environment eventually end up as Hot Jupiters, causing thus the high mean metallicity.

For a third population with $\f1=0.001$, but only partially suppressed type II migration, one finds results for the metallicities that approximately lie between the two cases discussed here. This means that the metallicity distribution of Hot Jupiters is a measure of the efficiency of both type I and type II migration. In summary we see that the synthetic populations reproduces the general observed trend, but that the more specific results depend on uncertain model settings like the efficiency of migration.

\section{Planet frequency}\label{sect:detectprob}
Comparing observed and theoretical detection probabilities  $P_{\rm det}$ of planets as a function of some  input (disk) variable is a classical application of population synthesis calculations (e.g. Ida \& Lin \cite{idalin2004b}; Kornet et al. \cite{kornetetal2005}; Dodson-Robinson et al. \cite{dodsonrobinsonetal2006}; Matsuo et al. \cite{matsuoetal2007}). Here we study the frequency of giant planets as a function of [Fe/H], $\mdisk$ and $\tdisk$.

\subsection{[Fe/H]: Influence of the gas mass}\label{sect:detectprobmetallicity}
In Paper II, we compared the synthetic ``metallicity effect'' (the increase of the detection rate of giant planets with host star [Fe/H]) to the observed one and found that the synthesis can reproduce this important observational constraint. Here we focus on another aspect, namely how the ``metallicity effect''  depends on the amount of gas present in a protoplanetary disk. 

For simplicity, we here again assume that a planet can be detected by the radial velocity method if it induces a velocity semi-amplitude $K$ larger than 10 m/s, and if its orbital period is less than 10 years. We thus focus on giant planets. 

Figure \ref{fig:pdetecfeh}, left panel, shows the detection probability as a function of [Fe/H]  if additionally the gas mass $\sigmanorm$ of the corresponding disk is either larger than some lower limit (blue dashed lines), or smaller than some upper limit (red dotted lines). In all cases, the detection probability  $P_{\rm det}$([Fe/H],$\sigmanorm$) obtained in this way was divided by the detection probability at a given [Fe/H] but including all $\sigmanorm$, i.e. the  (usual) detection probability as a function of [Fe/H] alone. This quantity is shown in the right panel of Fig. \ref{fig:pdetecfeh}. Like that, the differential influence of the gas mass on $P_{\rm det}$  is shown.

One notes that for disks which have a $\sigmanorm$ above a certain threshold (blue lines), i.e. where the most gas poor disks are excluded, the normalized probability is larger than 1, which means that more planets are detected. When only a moderate lower limit is used ($\sigmanorm>100$ g/cm$^{2}$), the increase is only small and nearly identical for all [Fe/H]. However, the higher the $\sigmanorm$ threshold becomes, the more a dependence on [Fe/H] develops: The relative increase of  $P_{\rm det}$ is highest for the subsolar metallicities. In other words, the metallicity effect is weakened, if the protoplanetary disks are gas rich. A mirrored effect is seen if only disks with a mass smaller than some $\sigmanorm$ (red lines) are considered, so that the most gas rich disks are now excluded: Here, the relative decrease of  $P_{\rm det}$ is strongest for the low [Fe/H] bins, which means that  the metallicity  is strengthened if the protoplanetary disks are gas poor. 

These findings are clearly a consequence of the ``compensation'' effect discussed theoretically in Sect. \ref{sect:compensationeffects}: In a gas rich environment, a lower [Fe/H] is sufficient to allow giant planet formation. 

\subsection{Disk gas mass}
\begin{figure}
\includegraphics[width=\columnwidth,angle=0]{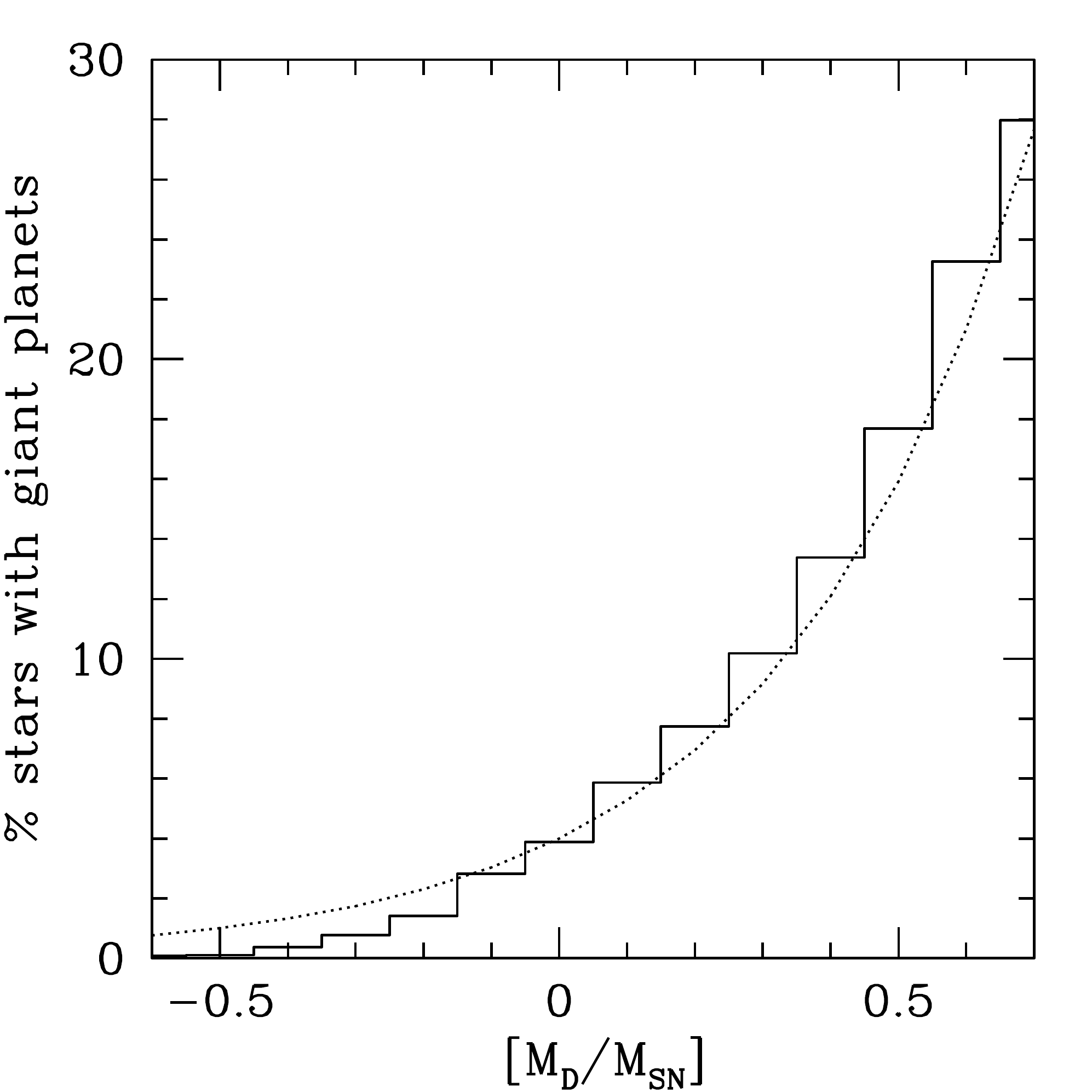}
\caption{Percentage of stars having a giant planet with a mass of at least $100\,\mearth$ as a function of the initial disk gas mass. The dotted line scales as $\mdisk^{1.2}$.  }\label{fig:pdetecsigma}
\end{figure}

Figure \ref{fig:pdetecsigma} shows the percentage of stars (initial conditions) where the formation of a giant planet with a mass of at least $100\,\mearth$ is possible, as a function of the initial disk gas mass measured as $\mm$. The lowest value of about -0.6 ($\sigmanorm=50$ g/cm$^{2}$) corresponds to an initial content of gas within the computational disk which of about 0.004 $\msun$. The upper limit of about 0.7  ($\sigmanorm=1000$ g/cm$^{2}$) corresponds to a disk with an initial mass of 0.09 $\msun$. As the individual primordial disk mass is unknown for the Gyr old stars around which extrasolar planets are typically discovered, we directly plot the real frequency, without taking into account any observational bias, in contrast to the results for [Fe/H].   

As expected from sections \ref{subsubsect:imf}, one sees that the frequency of giant planet is increasing with $\mm$. For $\mm\gtrsim-0.1$, the increase is found to be slightly stronger than linear with the disk mass, as indicated by the dotted fitting line which correspond to a fraction of stars with giant planets given as  $0.04\times(\mdisk/0.017 \msun)^{1.2}$. For lower disk masses, this fit overestimates the number of giant planets. We note that two quantities which are important for giant planet formation scale with similar exponents: the core accretion rate scales linearly with the planetesimal surface density $\sigmas$ which is itself proportional to the disk gas mass ($\sigmas=\fpg\sigmanorm)$, while the isolation mass is proportional to $\sigmas^{1.5}$ (e.g. Lissauer et al. \cite{lissauer1993}). 

For the most massive disks considered in this work, one finds a frequency of about 28\%. This is in fair agreement with the result of $\sim35\%$  of Dodson-Robinson et al. (\cite{dodsonrobinsonetal2006}) at the same disk mass. At lower disk masses we find somewhat higher frequencies than these authors. The reason for this difference could be that in Dodson-Robinson et al. (\cite{dodsonrobinsonetal2006}), migration was not included, so that higher surface densities are necessary (cf. sect. \ref{sect:minimalssd}).

\subsection{Disk lifetime}\label{sect:detectprobtdisk}
\begin{figure}
\includegraphics[width=\columnwidth,angle=0]{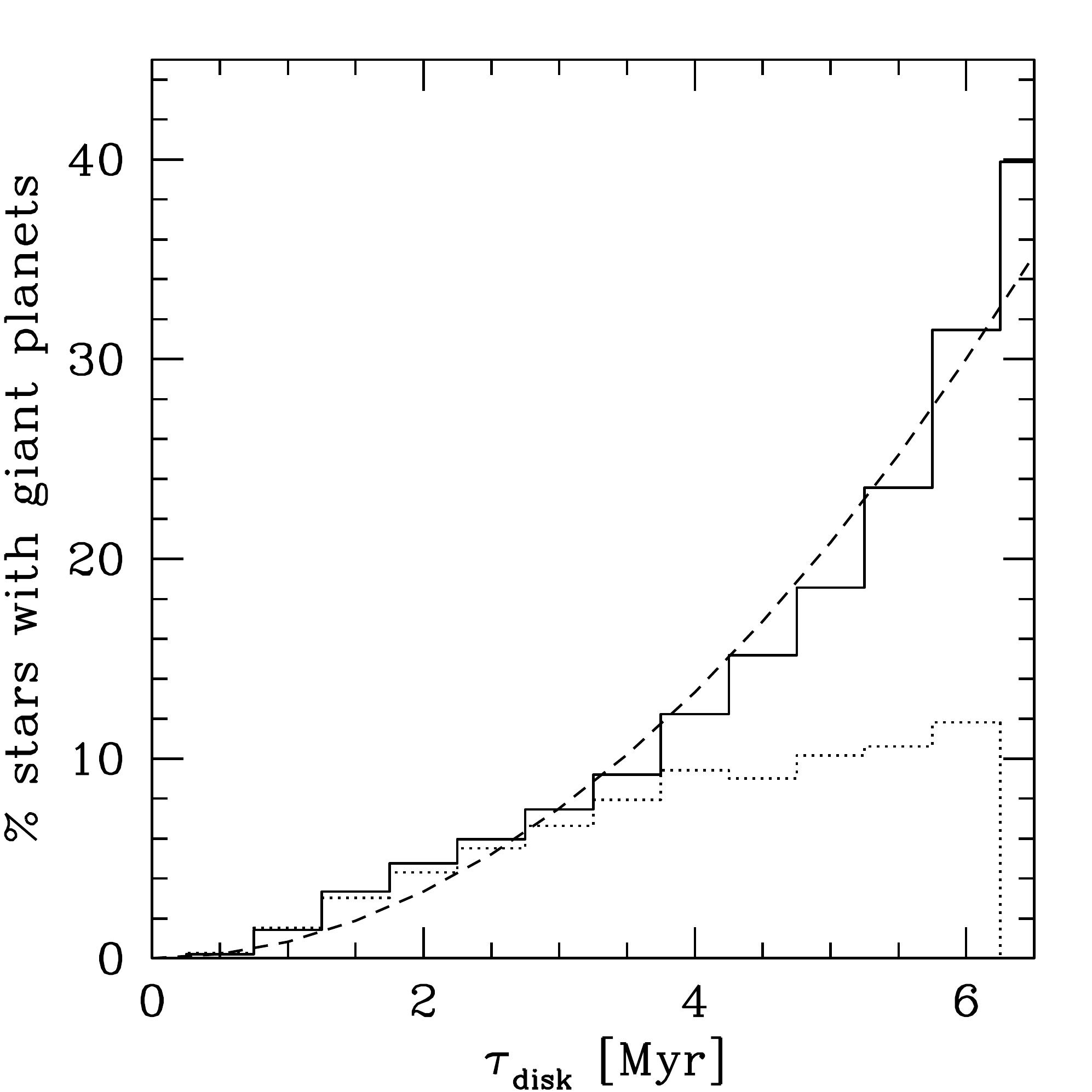}
\caption{Percentage of stars having a giant planet with a mass of at least $100\,\mearth$ as a function of the disk lifetime $\tdisk$.  The solid line shows the nominal population, while the dotted line corresponds to a population where disk mass and disk lifetime are not correlated (cf. sect. \ref{sect:influencesdisklifetime}).  The dashed lines scales as $\tdisk^{2}$.}\label{fig:pdetectdisk}
\end{figure}

Figure \ref{fig:pdetectdisk} shows the percentage of stars with a giant planet ($M\geq 100\,\mearth$) as a function of disk lifetime $\tdisk$. The solid line shows the dependence for the nominal population. Not surprisingly, there is a positive correlation of the disk lifetime and giant planet occurrence, as in longer living disk, cores have more time to grow to a size where they can trigger gas runaway accretion. The dashed lines shows an approximate fit to the frequency, given as $0.075\times (\tdisk/3\, \mathrm{Myr})^{2}$.  The strong dependence on $\tdisk$ is caused by the double role of $\tdisk$ both as a threshold, and as a factor directly influencing the planet mass, as described in Sect. \ref{subsubsect:imf}. 

For the disk with the longest $\tdisk\approx 6.5$ Myrs, the percentage of stars with a giant planet is remarkably high, namely about 40\%. For comparison, Dodson-Robinson et al. (\cite{dodsonrobinsonetal2006}) find about 30\% for such a $\tdisk$.  This high fraction is however dependent on a disk mass-disk lifetime correlation inherent in the initial conditions, as addressed in Section  \ref{sect:influencesdisklifetime} below.

Recently, it was discussed whether extrasolar planet host stars show a stronger depletion in lithium compared to non-planet hosts. No general agreement exists currently on this question on observational bases, with analyses both in favor (Israelian et al., \cite{israelianetal2009}, Sousa et al. \cite{sousaetal2010}) or against (Baumann et al. \cite{baumannetal2010}; Ghezzi et al. \cite{ghezzietal2010b}) such a correlation.  On theoretical bases, several mechanism can enhance  lithium depletion (e.g. Baraffe \& Chabrier \cite{baraffechabrier2010}). A possible explanation based on the star rotational history which is interesting in the context here has been put forward by Bouvier (\cite{bouvier2008}):  In this model it is assumed that long disk lifetimes lead via disc locking to a slower initial rotation of the star at the ZAMS.  This causes a phase of stronger differential rotation between the radiative core and the convective envelope, which is in turn responsible for an enhanced lithium depletion. 

If the correlation of enhanced lithium depletion and the presence of massive planets can indeed be confirmed, then the lines of reasoning of  Bouvier (\cite{bouvier2008}) provide a way of relating disk lifetime and the formation probability of giant planets. The result shown here that giant planets preferentially form around disks with a long lifetime fits well in this picture. Note that in the direct gravitational collapse model for giant planet formation, this is not  the case.  

The abundance of lithium would then provide a link which is observable today between primordial disk properties ($\tdisk$) and planet occurrence in similar, although much more indirect way than [Fe/H].  We comment that long disk lifetimes are in particular needed for giant planet formation when the metallicity is on the low side ([Fe/H]$\lesssim-0.2$), while at high [Fe/H], short lifetimes ($\tdisk\lesssim$ 2 Myrs) can be sufficient, as demonstrated in Section \ref{sect:longnecessarytdisk}, so that the planet frequency-$\tdisk$ correlation also depends on [Fe/H]. 


\section{Disk mass - disk lifetime correlation}\label{sect:influencesdisklifetime}
When generating initial conditions, we first draw in the nominal procedure the gas surface density $\sigmanorm$ and then, independently of it, the photoevaporation rate $\mwind$. This procedure has the side effect that disks with a high initial gas surface density $\sigmanorm$ will on average also have a higher lifetime $\tdisk$, as for the same $\mwind$  more massive disks life longer. Thus, there is a positive correlation between disk mass and lifetime. 

To assess the consequences of this correlation, we have generated an alternative set of non-nominal initial conditions in the following way: After drawing $\sigmanorm$, we directly draw a  $\tdisk$ from a specified distribution and then choose the $\mwind$ that is necessary to get this $\tdisk$, for the given $\sigmanorm$ (and the fixed $\alpha$). Like that, no correlations exist any more between $\sigmanorm$ and $\tdisk$.  Instead, $\sigmanorm$ and $\mwind$ are now positively correlated, which could arise in a situation where photoevaporation by the central star is dominant, even though that then, strictly speaking, the functional form of the photoevaporation would be different than assumed in our disk model (Matsuhama et al. \cite{matsuhamaetal2003}). 

The distributions we have used for $\tdisk$ for the non-nominal case are either a uniform distribution of lifetimes between 0 and 6 Myrs which corresponds to the linearly decreasing fraction of stars which have a JHKL excess  as a function of mean cluster age in Haisch et al. (\cite{haischetal2001}), or  an exponential decrease of this fraction with time as $\exp(-t/2.5 \ \rm Myr)$, which is  based on more recent compilations of disk lifetimes as in Mamajek (\cite{mamajek2009}) or Fedele et al. (\cite{fedeleetal2010}).

\subsection{Effect on the giant planet frequency} 
We have found that the impact of using this different prescription only has minor impact on the general properties of the synthetic population as compared to the nominal case. This regards in particular the distribution in the mass-distance plane, the migration behavior or the general shape of the PIMF. The difference between the linear and the exponential distribution for $\tdisk$ is even smaller. 

Only when we directly plot the frequency of planets as a function of disk lifetime as in Fig. \ref{fig:pdetectdisk}, a significant difference becomes visible. In this figure, the percentage of stars with a giant planet for the non-nominal case with the linear distribution is shown as the dotted line. One notes that while for the nominal case (solid line), a clear increase of the giant planet frequency with disk lifetime is seen for $\tdisk\gtrsim 4$ Myrs, this is not the case for the population without a correlation between $\tdisk$ and $\mdisk$. The reason for this is that in the non-nominal case, there are many disks with a long lifetime, but still only such a small $\sigmanorm$ that giant planets cannot form, which reduces the fraction. Before, disks with a high $\tdisk$ usually also had a high $\sigmanorm$, which contribute both to successful giant planet formation.  

\section{Summary and conclusion}\label{summaryconclusion}
We have extensively studied for solar-like stars the influence of some of the most important properties of their protoplanetary disks (disk metallicity [Fe/H], disk (gas) mass $\mdisk$ and lifetime $\tdisk$) on the planets growing in them. We have found numerous correlations which we summarize here. 

We first give a list of the most central correlations between disk and planetary properties as found within the core  accretion paradigm, which depend less sensitively on  model settings. From an observational point of view, we have to distinguish two classes: First correlations which can be tested against observations, as they involve the (stellar) [Fe/H]. Second correlations which are linked to the disk  mass and lifetime. These primordial disk properties cannot be directly observed, even though that some pathways to at least approximately determine them might arise in future. The correlations involving the metallicity are:
 \begin{enumerate}    
\item The planetary initial mass function for metal rich disks contains a higher number of giant planets, but not of a significantly higher mass. Metallicity mostly acts as a threshold for giant planet formation, but is not correlated  with the mass of giant planets, except for rare special cases (point \ref{itemmmax}).
\item No clear metallicity effect (positive correlation of the detection rate of planets with stellar [Fe/H]) is found in the Neptunian mass domain. At even lower masses, an anti-correlaton could exist.
\item Observationally, at a low radial velocity precision i.e. for giant planets, no clear correlation of [Fe/H] and the planetary mass distribution is visible. At high precision ($\sim1$ m/s) in contrast, a systematic anti-correlation of [Fe/H] and the Neptunian to Jovian planet ratio becomes detectable: The lower [Fe/H], the higher the ratio of Neptunian to Jovian planets.
\item The most massive companions ($\gtrsim10-20\,\mj$) cannot form by core accretion at clearly subsolar [Fe/H], as core growth takes then so long that disks are significantly depleted once the cores trigger gas runaway accretion. In the supersolar regime, maximal planet masses are in contrast independent of [Fe/H]. Such planets are however anyway very rare. \label{itemmmax}
\item The absence of a strong correlation of metallicity and the semimajor axis distribution of giant planet is explained like this: Around metal poor stars, giant planet cores can only form in gas rich disk, and at large distances $\astart\gtrsim\aice$. These high gas disk masses cause migration over larger distances, tending to cancel out the large $\astart$. In metal rich disk it is the contrary: giant planets can form also at smaller distances. But as low gas masses are sufficient, planets migrate also less. These correlations of accretion and migration explain the observed weak dependence.  In short: at low [Fe/H], planets start further out, but migrate more, while at high [Fe/H] they start closer in, but migrate less. 
\end{enumerate}

The correlations involving the disk mass and  lifetime are:
\begin{enumerate}    
\item The planetary initial mass function for massive gas disks contains more giants planets of a higher mass,  but less giants of a lower mass. Disk gas masses and giant planet masses are correlated.
\item Disk lifetimes act both as a threshold (like [Fe/H]) for giant planet formation but also influence the final total mass (like $\mdisk$), so that they have a strong influence on the giant planet population. 
\item Maximal planetary masses and disk mass are in contrast to the situation for [Fe/H] correlated over the full range of disk masses. Except for very low disk masses, maximal possible planet mass and disk mass correlate linearly. There is a wide spread in the efficiencies of converting disk gas in envelope mass, but to order of magnitude it is 10\%.
\item Low metallicities (i.e. dust-to-gas ratios) can be compensated by high disk masses to allow giant planet formation, and vice versa (``compensation effect''). The lowest necessary total solid content in a disk for giant planet formation is about 0.4 to 0.5 $\mj$. 
\item The metallicity effect is weaker when disk gas masses are high, and stronger if they are small. This is a consequence of the metallicity-disk gas mass ``compensation effect''. 
\item Near the solid surface density threshold necessary for giant planet formation, long disk lifetimes of at least 3-4 Myrs are necessary. This could indicate that the Solar Nebula probably had a rather long lifetime. 
\end{enumerate}

We have also found the following correlations, which are either of second order, or could be dependent in a more significant way on specific model assumptions:
\begin{enumerate}
\item There is an imprint of the disk mass distribution on the upper end of the planetary mass function. 
\item The chain of correlations of low [Fe/H]$ \Rightarrow$ high necessary gas mass $\Rightarrow$ large $\aice \Rightarrow$ large typical $\astart$ means that in low [Fe/H] disks, embryos of giant planets come typically from further out. The opposite applies for high [Fe/H].
\item For giant planet formation, low metallicities can not be compensated by high gas masses ad infinitum, at least if higher mass disks have an ice line further out due to stronger viscous dissipation: The sweet spot for giant planet formation at $\astart$$\sim$$\aice$ moves then so far out that the increasing core growth timescale can no more be compensated for by increasing the disk mass. 
\item The semimajor axis distribution of giant planets outside 0.1 AU first consists of a nearly flat part, followed by an upturn. The location of the upturn probably depends on the location of the ice line. It is due to the fact that there is a typical locus from where giant planets come, combined with a typical distance over which they migrate. Disk thermodynamical properties thus leave an imprint in the planetary semimajor axis distribution. 
\item The extent of migration of giant planets $\triangle a =a-\astart$ is large ($\sim-6$ AU)  in gas rich disks with a low [Fe/H]: Torques are strong, $\astart$ are large, and cores must migrate over a large distance until they have collected sufficient planetesimals to trigger gas runaway accretion and slow down in slow planet dominated type II migration. The extent is minimal ($\lesssim-1$ AU) in gas and solid rich disk and if the embryo starts inside $\aice$: giant planets grow rapidly and local disk masses are  small, so that planets soon stop due to their inertia. For low mass disks $\triangle a\sim$-3 AU, and for disks similar to the MMSN,  $\triangle a\sim$-4 AU. 
\item The extent of migration of giant planets increases with disk lifetime $\tdisk$. Very roughly, there is a linear correlation. For disks with the shortest $\tdisk\approx 0.5$ Myrs still allowing giant planet formation, $\triangle a$ is  $\sim$ -2 AU, while in the longest living disks ($\tdisk\approx7$ Myrs),   $\triangle a \sim$ -5.5 AU. There are important systematic departures from this correlation depending on the solid content of the disk.
\item Observationally, Hot Jupiters are found around higher metallicity hosts than lower mass close-in Neptunian and Super Earth planets. This is also found in the synthetic population. The degree of difference correlates positively with the efficiency of type I migration and varies between 0.03 and 0.11 dex for the mean [Fe/H] values.   
\end{enumerate}

Finally, with the following points we mostly confirm earlier results concerning the relationship of disk properties and planet properties. The quantitative results often differ to a certain degree, but  qualitatively they are the same (cf. Ida \& Lin \cite{idalin2004a,idalin2004b};  Dodson-Robinson et al. \cite{dodsonrobinsonetal2006}; Kornet et al. \cite{kornetetal2006};  Matsuo et al. \cite{matsuoetal2007};  Thommes et al. \cite{thommesetal2008}).
 \begin{enumerate}
\item Comparison with observation shows that the very large majority of known extrasolar planets lies in the [Fe/H]-mass domain populated by the core accretion formation mechanism.
\item  The sweet spot for planet formation is at about 5 AU. The minimal necessary surface density of planetesimals to allow giant planet formation is about $\sigmas=6$ g/cm$^{2}$ there. Both in- and outside this distance, higher $\sigmas$ are necessary. Inside, to overcome the small isolation mass, outside to overcome the long core growth timescale. 
\item At metallicities [Fe/H]$\lesssim-0.1$, giant planets can only form in an annulus of 2-3 AU in width outside the ice line $\aice$. With increasing [Fe/H], giant planets can also form in- and clearly outside $\aice$, but the ice line remains the preferred starting position.  Only at the highest [Fe/H]$\gtrsim0.4$, the location of $\aice$ becomes less important and giant planets can form all over the disk from $\sim1$ to $\sim20$ AU. 
\item The fraction of stars with a giant planet as a function of initial disk mass is found to be approximately $0.04\times(\mdisk/0.017 \msun)^{1.2}$.
\item The fraction of stars with a giant planet as a function of disk lifetime is found to be roughly  $0.075\times(\tdisk/3\,\mathrm{Myr})^{2}$. A possible link between giant planet occurrence and lithium depletion in the host star could be an observational manifestation of this correlation.  The influence of $\tdisk$ on the frequency of giant planets is however much reduced if the disk mass and lifetime are not positively correlated, as implicitly assumed in the nominal population.   
\end{enumerate}



This long list of correlations clearly indicates the strong dependence of planet formation on protoplanetary disk properties.  It calls for profound improvements of the disk models used in planet formation simulations, for a rigid inclusion of observational results on disks in the initial conditions, and for the incorporation of observations of disks like SEDs as additional constraints for planet formation models.  

Such improvements of the initial and boundary conditions for planet formation will in the end allow to better understand the planet formation process itself. 

\textit{Note:} The numerical data of the nominal synthetic population can be obtained in electronic form at \href{http://www.mpia.de/homes/mordasini/Site7.html}{http://www.mpia.de/homes/mordasini/Site7.html}

\acknowledgements
We thank Stephane Udry, Chris Ormel and Kai-Martin Dittkrist for useful discussions. C. Mordasini acknowledges the financial support as a fellow of the Alexander von Humboldt foundation. This work was supported in part by the Swiss National Science Foundation.  Yann Alibert is thankful for the  support by the ERC under grant 239605. C. Mordasini acknowledges the hospitality of the Kavli Institute at UCSB in spring 2010 (U. S. NSF grant PHY05-51164). Calculations were made on PANSTARRS at the MPG Rechenzentrum Garching. We thank an anonymous referee for helpful suggestions.

\listofobjects
\end{document}